\documentclass[%
preprint,
amsmath,amssymb,
aps,
prfluids,
groupedaddress,
]{revtex4-2}

\usepackage{graphicx}
\usepackage{dcolumn}
\usepackage{bm}
\usepackage{lipsum} 
\usepackage{float}
\usepackage{mathrsfs,amsthm,amsmath,amssymb}
\usepackage[colorlinks=true,linkcolor=blue]{hyperref}%
\usepackage{color}

\begin{document}


\title{Prediction of turbulent channel flow using Fourier neural operator-based machine-learning strategy}
\author{Yunpeng Wang$^{1,2,3}$}
\author{Zhijie Li$^{1,2,3}$}
\author{Zelong Yuan$^{1,2,3,4}$}
\author{Wenhui Peng$^{5}$}
\author{Tianyuan Liu$^{6}$}
\author{Jianchun Wang$^{1,2,3}$}
\email{wangjc@sustech.edu.cn}
\affiliation{\small 1. Department of Mechanics and Aerospace Engineering, Southern University of Science and Technology, Shenzhen, 518055 China.\\
2. Guangdong Provincial Key Laboratory of Turbulence Research and Applications, Southern University of Science and Technology, Shenzhen 518055,China.\\
3. Guangdong-Hong Kong-Macao Joint Laboratory for Data-Driven Fluid Mechanics and Engineering Applications, Southern University of Science and Technology, Shenzhen, 518055 China.\\
4. Harbin Engineering University Qingdao Innovation and Development Base, Qingdao 266000, China.\\
5. The Hong Kong Polytechnic University, Department of Applied Mathematics, 999077, Hong Kong Special Administrative Region of China.\\
6. Engineering College, Peking University, Beijing, 100091, China.}
\date{\today}

\begin{abstract}
Fast and accurate predictions of turbulent flows are of great importance in the science and engineering field. In this paper, we investigate the implicit U-Net enhanced Fourier neural operator (IUFNO) in the stable prediction of long-time dynamics of three-dimensional (3D) turbulent channel flows. The trained IUFNO models are tested in the large-eddy simulations (LES) at coarse grids for three friction Reynolds numbers: $Re_{\tau}\approx180$, $395$ and $590$. The adopted near-wall mesh grids are tangibly coarser than the general requirements for wall-resolved LES. Compared to the original Fourier neural operator (FNO), the implicit FNO (IFNO) and U-Net enhanced FNO (UFNO), the IUFNO model has a much better long-term predictive ability. The numerical experiments show that the IUFNO framework outperforms the traditional dynamic Smagorinsky model (DSM) and the wall-adapted local eddy-viscosity (WALE) model in the predictions of a variety of flow statistics and structures, including the mean and fluctuating velocities, the probability density functions (PDFs) and joint PDF of velocity fluctuations, the Reynolds stress profile, the kinetic energy spectrum, and the Q-criterion (vortex structures). Meanwhile, the trained IUFNO models are computationally much faster than the traditional LES models. Thus, the IUFNO model is a promising approach for the fast prediction of wall-bounded turbulent flow.

\end{abstract}

\pacs{Valid PACS appear here}
\maketitle

\section{\label{sec:level1}Introduction}

Turbulent flows are ubiquitous in meteorology, aerospace engineering, air pollution control, geosciences and industrial activities \cite{Pope2000}. Among many flow prediction methods, computational fluid dynamics (CFD) is an important tool as it can provide useful estimations of the flow field especially when experimental measurements are challenging \cite{Bauweraerts2021}. Even so, due to the large range of motion scales involved, the direct numerical simulation (DNS) of turbulence is still impractical at high Reynolds numbers \cite{Meneveau2000,Yang2021,Moser2021}. As a result, coarse-grid simulations are often adopted including the Reynolds-averaged Navier-Stokes (RANS) method and the large-eddy simulation (LES). RANS method only solves the mean flow field and has been widely adopted for industrial applications \cite{Durbin2018,Pope1975,Duraisamy2019,Jiang2021,Chen2023,Bin2024}. On the other hand, LES can directly resolve the major energy-containing large-scale motions, and hence, can predict the flow structures better \cite{Sagaut2006,Moin1991,Smagorinsky1963,Lilly1967,Deardorff1970,Germano1992,Mons2021,Rouhi2016,Celora2021,Rozema2022}. However, the computation cost for LES is higher than that for RANS, and it is even close to the computation cost for DNS in the case of wall-resolved LES \cite{Yang2021}.

In the recent decade, machine-learning-based flow prediction methods have been extensively proposed due to the fast development of modern computers and the accumulation of high fidelity data\cite{Park2021,Novati2021,Bae2022,Guan2022,Wu2024a,Li2024}. These attempts include, but not limited to, learning the important aerodynamic forces in the flow field such as the lifting and dragging force coefficient \cite{Wang2022,Mohamed2023,Kurtulus2009}, reconstructing part of governing equations such as the modeling of the RANS and LES closure terms \cite{Ling2016,Wang2017,Wu2018,Wu2019,Yang2019,Gamahara2017,Maulik2017,Beck2019,Maulik2019,Zhou2019,Xie2019,Xie2019a,Xie2020,Yuan2021,Kurz2023}, wall modeling \cite{LozanoDuran2023,Vadrot2023}, inferring the missing information in the case of measurement constraints or flow field damage \cite{Buzzicotti2021,Guastoni2021,Li2023a,Zhu2024}, flow field super-resolution \cite{Deng2019,Kim2021}, and directly predicting the evolution of the flow field \cite{Deng2023,Li2020,Li2022a,Peng2022,Peng2023,Mohan2019,Mohan2020,Nakamura2021,Bukka2021,Srinivasan2019,Li2023b,Lienen2023,Wu2024,Gao2024,Du2024,Kohl2024}. Among these applications, directly predicting the temporal evolution of the flow field has attracted increasing attentions these days, since the trained model does not require to solve the Navier-stokes equations while giving a fast evaluation of the detailed flow information. The adopted methods mainly include the recursive neural network (RNN) and long short term memory (LSTM)-based frameworks \cite{Mohan2019,Mohan2020,Han2019,Han2021,Han2022,Nakamura2021,Bukka2021}, physics-informed neural networks (PINN)-based methods \cite{Raissi2019,Wang2020,Jin2021}, and neural operator-based methods \cite{Deng2023,Li2020,Li2022,Peng2022,Peng2023}. For example, Bukka et al. combined the proper orthogonal decomposition (POD) and deep learning approach to predict the flow past a cylinder \cite{Bukka2021}. Han et al. conducted a series of studies using the CNN and LSTM-based framework in the predictions of flow field and fluid-solid interactions \cite{Han2019,Han2021,Han2022}. Raissi et al. developed a PINN-based framework to predict the solutions of general nonlinear partial differential equations \cite{Raissi2019}. In the two-dimensional (2D) Rayleigh Bénard convection problem, Wang et al. developed a turbulent flow network (TF-Net) based on the specialized U-Net with incorporated physical constraints \cite{Wang2020}.

While many neural networks (NN) are effective at approximating the mappings between finite-dimensional Euclidean spaces for a given data set, these NN-based models are difficult to generalize to different flow conditions or boundary conditions \cite{Raissi2019,Li2020}. In this consideration, Li et al. proposed a novel Fourier neural operator (FNO) framework which can efficiently learn the mappings between the high dimensional features in Fourier space, and it enables reconstructing the information in infinite dimensional spaces \cite{Li2020}. In their numerical tests, the proposed FNO model achieves outstanding accuracy in the prediction of two-dimensional (2D) turbulence. Since Li et al., many extensions and applications of FNO have emerged \cite{Li2022,Li2022a,Li2023,Lehmann2023,Peng2022,Peng2023,Wen2022,Deng2023,Lyu2023,Kurth2023,Li2023c,Peng2024,Meng2023,You2022,Qin2024,Qi2024,Li2024a,Li2023d}. Lehmann et al. applied the FNO in the prediction of the propagation of seismic waves \cite{Lehmann2023}. To improve the prediction accuracy of FNO for 2D turbulence at high Reynolds numbers, Peng et al. introduced the attention mechanism into the FNO framework, resulting in better statistical properties and instantaneous structures of the flow field \cite{Peng2022}. Wen et al. developed a U-Net enhanced FNO (UFNO) to solve the complex gas-liquid multiphase problems with improved accuracy compared to the original FNO and the CNN frameworks \cite{Wen2022}. You et al. proposed an implicit Fourier neural operator (IFNO) \cite{You2022}, which can greatly reduce the number of trainable parameters and memory cost compared to multi-layer structure of the FNO \cite{Li2023}.

Even though many NN-based flow prediction methods have been proposed, we should note that most of these methods are developed for laminar flows or 2D turbulent problems. In comparison, the NN-based predictions for 3D turbulence are investigated to a much lesser extent. The nonlinear interactions in 3D turbulence are fundamentally more complex than 2D turbulence. Consequently, modeling 3D turbulence is more challenging, considering the increased model complexity, memory usage and number of NN  parameters \cite{Li2022}. In the prediction of 3D homogeneous isotropic turbulence and scalar turbulence, Mohan et al. proposed two reduced models based on the convolutional generative adversarial network (C-GAN) and compressed convolutional long-short-term-memory (CC-LSTM) network \cite{Mohan2019,Mohan2020}. The reconstructed statistics are close to the DNS results. Peng et al. proposed a linear attention mechanism-based FNO (LAFNO) to predict the 3D homogeneous isotropic turbulence and turbulent mixing layer with improved accuracy and efficiency \cite{Peng2023}. Recently, to enable the modeling of 3D turbulence at high Reynolds numbers, Li et al. have trained the FNO and implicit U-Net enhanced FNO (IUFNO) using the coarse-grid filtered DNS (fDNS) data of 3D isotropic turbulence and turbulent mixing layer \cite{Li2022,Li2023}. In their implementation, the obtained NN model can be viewed as a surrogate model for LES, and it can predict the long-term dynamics of 3D turbulence with adequate accuracy and stability \cite{Li2023}. While these NN models are certainly useful in the prediction of 3D turbulence, most of them are limited to the unbounded flow situations (i.e. in the absence of solid wall).

For wall-bounded flows, Nakamura et al. combined a 3D CNN autoencoder (CNN-AE) and a LSTM network to predict the 3D turbulent channel flow \cite{Nakamura2021}. The model is tested at low friction Reynolds number ($Re_{\tau}\approx110$), and the predicted statistics agree with the DNS results. Using the PINN framework, Jin et al. developed a Navier-Stokes flow nets (NSFnets) \cite{Jin2021}, which can be applied to turbulent channel flows. In their work, the PINN-based solutions can be obtained in small subdomains, given the initial and boundary conditions of the subdomain. To date, developing a NN model to predict all scales of turbulence (i.e. a surrogate model for DNS) for wall-bound 3D turbulence at moderately high Reynolds numbers is yet challenging due to the potentially huge number of NN parameters. Hence, we aim at predicting the dominant energy containing scales, i.e., constructing a surrogate model for LES of wall-bounded 3D turbulent flow. In the current work, the FNO-based LES strategies are explored and tested for 3D turbulent channel flows at three friction Reynolds numbers $Re_{\tau}\approx180$, $Re_{\tau}\approx395$ and $Re_{\tau}\approx590$. To the best of our knowledge, this is the first attempt to construct surrogate LES models for wall-bounded 3D turbulent flows at moderately high Reynolds numbers. As shall be seen, the present model is very promising compared to the traditional LES models.

The rest of the paper is organized as follows. The governing equations of incompressible turbulence and traditional LES are introduced in Section II, followed by a brief discussion on the solution strategies for the LES equations and their respective advantages and shortcomings. The FNO and implicit U-Net enhanced FNO (IUFNO) frameworks are introduced in Section III. In Section IV, the performances of the FNO and IUFNO frameworks are evaluated in the LESs of turbulent channel flows at different friction Reynolds numbers. Finally, a brief summary of the paper and some future perspectives are given in Section V.

\section{Governing equations of incompressible turbulence and the large-eddy simulation}

The governing equations of incompressible turbulence are first introduced in this section, followed a brief description of the traditional LES strategy and some classical LES models.

For an incompressible Newtonian fluid, the mass and momentum conservation are governed by the 3D Navier-Stokes (NS) equations, namely \cite{OConnor2024,Ishihara2009}:
 \begin{equation}
  \frac{\partial u_{i}}{\partial x_{i}}=0,
  \label{mass}
\end{equation}
 \begin{equation}
  \frac{\partial u_{i}}{\partial t}+\frac{\partial u_{i} u_{j}}{\partial x_{j}}=-\frac{\partial p}{\partial x_{i}}+\nu\frac{\partial^{2} u_{i}}{\partial x_{j}\partial x_{j}}+\mathcal{F}_{i},
  \label{momentum}
\end{equation}
where $u_{i}$ is the $i$th velocity component, $\nu$ is the kinematic viscosity, $p$ is the pressure divided by the constant density $\rho$, and $\mathcal{F}_i$ accounts for any external forcing. Throughout this paper, the summation convention is used unless otherwise specified. For wall-bounded turbulence, the friction Reynolds number is defined as

 \begin{equation}
  Re_{\tau}=\frac{u_{\tau}\delta_{\nu}}{\nu},
  \label{Re_tau}
\end{equation}
where $\delta_{\nu}=\nu/u_{\tau}$ is viscous length scale and $u_{\tau}=\sqrt{\tau_{w}/\rho}$ is the wall shear velocity. Here the wall-shear stress is calculated as $\tau_{w}=\mu \partial \langle u \rangle/\partial y$ at the wall ($y=0$), with $\langle \cdot \rangle$ denoting a spatial average over the homogeneous streamwise and spanwise directions.

Even though the NS equations have be discovered for more than a century, seeking the full-scale solutions of these equations using DNS is yet impractical at high Reynolds numbers \cite{Pope2000,Meneveau2000,Yang2021,Moser2021}. Unlike DNS, LES only solves the major energy-containing large-scale motions using a coarse grid, leaving the subgrid motions handled by the SGS models \cite{Sagaut2006,Moin1991,Germano1992,Smagorinsky1963,Lilly1967,Deardorff1970}. The governing equations for LES are obtained through a filtering operation as follows
\begin{equation}
  \overline{f}(\mathbf{x})=\int_{D}f(\mathbf{x}-\mathbf{r})G(\mathbf{r},\mathbf{x};\overline{\Delta})d\mathbf{r},
  \label{filtering}
\end{equation}
where $f$ can be any physical quantity of interest, G is the filter kernel, $\overline{\Delta}$ is the filter width and D is the physical domain. Applying Eq.~(\ref{filtering}) to Eqs.~(\ref{mass}) and (\ref{momentum}) yields \cite{Meneveau2000,Sagaut2006}
\begin{equation}
  \frac{\partial \overline{u}_{i}}{\partial x_{i}}=0,
  \label{fiteredmass}
\end{equation}
 \begin{equation}
  \frac{\partial \overline{u}_{i}}{\partial t}+\frac{\partial \overline{u}_{i} \overline{u}_{j}}{\partial x_{j}}=-\frac{\partial \overline{p}}{\partial x_{i}}-\frac{\partial\tau_{ij}}{\partial x_{j}}+\nu\frac{\partial^{2} \overline{u}_{i}}{\partial x_{j}\partial x_{j}}+\overline{\mathcal{F}}_{i},
  \label{filteredmomentum}
\end{equation}
where the unclosed SGS stress $\tau_{ij}$ is defined by
\begin{equation}
  \tau_{ij}=\overline{u_{i}u_{j}}-\overline{u}_{i}\overline{u}_{j},
  \label{tau}
\end{equation}
and represents the nonlinear interactions between the resolved and subgrid motions. Apparently, to solve the LES equations, the SGS stress must be modeled in terms of the resolved variables.

A very well-known SGS model is the Smagorinsky model (SM) \cite{Smagorinsky1963}, which can be written as

\begin{equation}
 \tau^{A}_{ij}=\tau_{ij}-\frac{\delta_{ij}}{3}\tau_{kk}=-2C^{2}_{Smag}\overline{\Delta}^{2}|\overline{S}|\overline{S}_{ij},
  \label{tauDSM}
\end{equation}
with $\overline{\Delta}$ being the filter width and $\overline{S}_{ij}$ the filtered strain rate. $|\overline{S}|=\sqrt{2\overline{S}_{ij} \overline{S}_{ij}}$ is the characteristic filtered strain rate. The classical value for the Smagorinsky coefficient is $C_{Smag}=0.16$, which can be determined through theoretical arguments for isotropic turbulence \cite{Smagorinsky1963,Lilly1967,Pope2000}.

The Smagorinsky model is known to be over-dissipative in the non-turbulent regime, and should be attenuated near walls \cite{Pope2000}. To resolve this issue, a dynamic version of the model, the dynamic Smagorinsky model (DSM), has been proposed \cite{Germano1992}. In the DSM, an appropriate local value of the Smagorinsky coefficient is determined using the Germano identity \cite{Germano1992,Pope2000,Moin1991,Lilly1992,Meneveau2000,Sagaut2006}. Through a least-square method, the coefficient $C^{2}_{Smag}$ can be calculated as
\begin{equation}
  C^{2}_{Smag}=\frac{\langle L_{ij} M_{ij}\rangle}{\langle M_{kl} M_{kl}\rangle},
  \label{C}
\end{equation}
where $L_{ij}=\widetilde{\overline{u}_{i}\overline{u}_{j}}-\widetilde{\overline{u}}_{i}\widetilde{\overline{u}}_{j}, \alpha_{ij}=-2\overline{\Delta}^{2}|\overline{S}|\overline{S}_{ij}, \beta_{ij}=-2\widetilde{\overline{\Delta}}^2|\widetilde{\overline{S}}|\widetilde{\overline{S}}_{ij}$ and $M_{ij}=\beta_{ij}-\widetilde{\alpha}_{ij}$. Here the overbar denotes the filtering at scale $\overline{\Delta}$, and a tilde denotes a coarser filtering ($\widetilde{\Delta}=2\overline{\Delta}$).

For wall-bounded turbulent flows, another widely used model is the wall-adapting local eddy-viscosity (WALE) model \cite{Nicoud1999}, which can recover well the near-wall scaling without any dynamic procedure. The WALE model can be written as

\begin{equation}
 \tau^{A}_{ij}=\tau_{ij}-\frac{\delta_{ij}}{3}\tau_{kk}=2\nu_t\overline{S}_{ij},
  \label{tauWALE}
\end{equation}
where

\begin{equation}
 \nu_t = (C_w \Delta)^2 \frac{(\mathcal{S}^d_{ij} \mathcal{S}^d_{ij})^{3/2}}{(\overline{S}_{ij} \overline{S}_{ij})^{5/2}+(\mathcal{S}^d_{ij} \mathcal{S}^d_{ij})^{5/4}}.
  \label{nu_t}
\end{equation}
Here, $C_w=0.5$, $\mathcal{S}^d_{ij} \mathcal{S}^d_{ij} = \frac{1}{6}(S^2S^2+\Omega^2\Omega^2)+\frac{2}{3}S^2\Omega^2+2IV_{S\Omega}$, where $S^2=\overline{S}_{ij} \overline{S}_{ij}$, $\Omega^2=\overline{\Omega}_{ij} \overline{\Omega}_{ij}$ and $IV_{S\Omega}=\overline{S}_{ik} \overline{S}_{kj} \overline{\Omega}_{jl} \overline{\Omega}_{li}$.

More detailed derivations of these models can be found in the literature and not reproduced \cite{Germano1992,Nicoud1999,Xie2020,Yuan2020}. In the current work, both the DSM and WALE models will be tested in the LES of turbulent channel flows.

\section{The Fourier neural operator and the implicit U-Net enhanced Fourier neural operator}

While many NN-based methods focus on reconstructing the nonlinear mappings of flow field in the physical domain, the FNO framework learns the mappings of the high-dimensional data in the frequency domain. In this case, the nonlinear operators can be approximated by the learned relationships between the Fourier coefficients. More importantly, FNO can truncate the less important high-frequency modes and only evolve the dominant large-scale modes \cite{Li2022}. In this section, the FNO and implicit U-Net enhanced FNO (IUFNO) are introduced.

\subsection{The Fourier neural operator}

\begin{figure}\centering
\includegraphics[width=.9\textwidth]{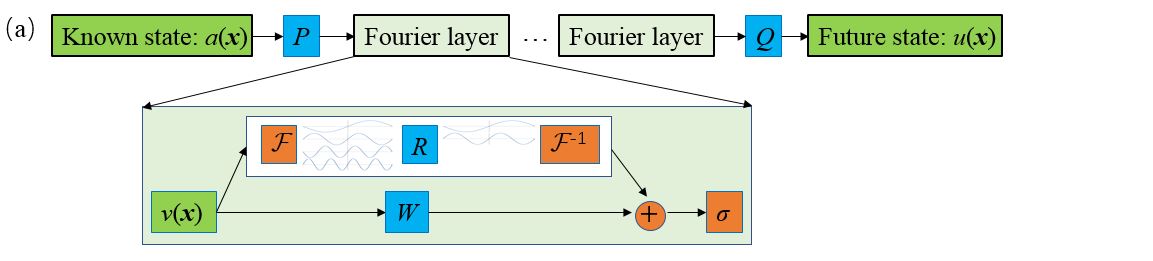}\hspace{-0.10in}

\includegraphics[width=.9\textwidth]{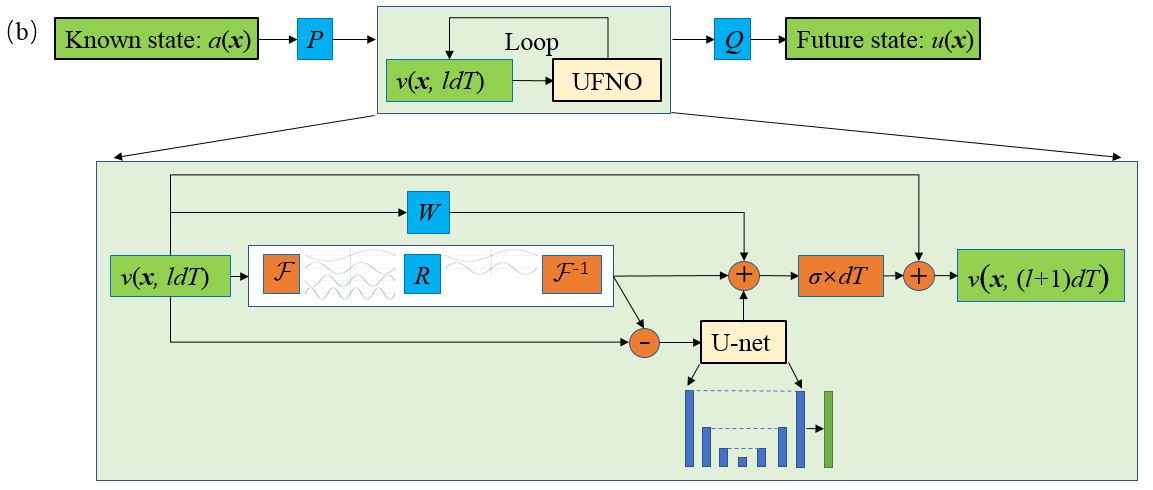}\hspace{-0.10in}
 \caption{The configurations of (a) Fourier neural operator (FNO); and (b) implicit U-Net enhanced Fourier neural operator (IUFNO).}\label{fig_config}
\end{figure}

Given a finite set of input-output pairs, the Fourier neural operators (FNO) aims to map between two infinite-dimensional spaces. Denote $D \subset \mathbb{R}^d$ as a bounded, open set and $\mathcal{A}=\mathcal{A}\left(D ; \mathbb{R}^{d_a}\right)$ and $\mathcal{U}=\mathcal{U}\left(D ; \mathbb{R}^{d_u}\right)$ as separable Banach spaces of function taking values in $\mathbb{R}^{d_a}$ and $\mathbb{R}^{d_u}$ respectively \cite{Beauzamy2011}. The construction of a mapping, parameterized by  $\theta \in \Theta$, allows FNO to learn an approximation of $\mathcal{A} \rightarrow \mathcal{U}$. The FNO architecture is shown in Fig.~\ref{fig_config}a, and described as follows:

(1) The input variables $a(\mathbf{x})$, being the known states in the current work, are projected to a higher dimensional representation $v(\mathbf{x})$ through the transformation $P$ parameterized by a shallow fully connected neural network.

(2) The higher dimensional variables $v(\mathbf{x})$, which take values in $\mathbb{R}^{d_v}$, are updated between the Fourier layers by
  \begin{equation}
  v_{m+1}(\mathbf{x})=\sigma\left(W v_m(\mathbf{x})+\left(\mathcal{K}(a ; \phi) v_m\right)(\mathbf{x})\right), \quad \forall \mathbf{x} \in D,  
  \label{v_update}
  \end{equation}
where $m$ denotes the $m$th Fourier layer, $\mathcal{K}: \mathcal{A} \times \Theta_{\mathcal{K}} \rightarrow \mathcal{L}\left(\mathcal{U}\left(D ; \mathbb{R}^{d_v}\right), \mathcal{U}\left(D ; \mathbb{R}^{d_v}\right)\right)$ maps to bounded linear operators on $\mathcal{U}\left(D ; \mathbb{R}^{d_v}\right)$ and is parameterized by $\phi \in \Theta_{\mathcal{K}}$, $W: \mathbb{R}^{d_v} \rightarrow \mathbb{R}^{d_v}$ is a linear transformation, and $\sigma: \mathbb{R} \rightarrow \mathbb{R}$ is non-linear local activation function.

(3) The output function $u \in \mathcal{U}$ is obtained by $u(\mathbf{x})=$ $Q\left(v_m(\mathbf{x})\right)$ where $Q: \mathbb{R}^{d_v} \rightarrow \mathbb{R}^{d_u}$ is the projection of $v_m$ and it is parameterized by a fully connected layer \cite{Li2020}.

By letting $\mathcal{F}$ and $\mathcal{F}^{-1}$ denote the Fourier transform and its inverse transform of a function $f: D \rightarrow \mathbb{R}^{d_v}$ respectively, and substituting the kernel integral operator in Eq.~\ref{v_update} with a convolution operator defined in Fourier space, the Fourier integral operator can be written as 
  \begin{equation}
  \left(\mathcal{K}(\phi) v_m\right)(\mathbf{x})=\mathcal{F}^{-1}\left(R_\phi \cdot\left(\mathcal{F} v_m\right)\right)(\mathbf{x}), \quad \forall \mathbf{x} \in D,
  \label{eq10}
  \end{equation}
where $R_\phi$ is the Fourier transform of a periodic function $\mathcal{K}: \bar{D} \rightarrow \mathbb{R}^{d_v \times d_v}$ parameterized by $\phi \in \Theta_{\mathcal{K}}$. The frequency mode $k \in \mathbb{Z}^d$. The finite-dimensional parameterization is obtained by truncating the Fourier series at a maximum number of modes $k_{\max }=\left|Z_{k_{\max }}\right|=\mid\left\{k \in \mathbb{Z}^d:\left|k_j\right| \leq k_{\max , j}\right.$, for $\left.j=1, \ldots, d\right\} \mid$. $\mathcal{F}\left(v_m\right) \in \mathbb{C}^{n \times d_v}$ can be obtained by discretizing domain $D$ with $n \in \mathbb{N}$ points, where $v_m \in \mathbb{R}^{n \times d_v}$ \cite{Li2020}. By simply truncating the higher modes, $\mathcal{F}\left(v_m\right) \in \mathbb{C}^{k_{\max } \times d_v}$ can be obtained, where $\mathbb{C}$ is the complex space. $R_\phi$ is parameterized as complex-valued-tensor $({k_{\max } \times d_v \times d_v})$ containing a collection of truncated Fourier modes $R_\phi \in \mathbb{C}^{k_{\max } \times d_v \times d_v}$. By multiplying $R_\phi$ and $\mathcal{F}\left(v_m\right)$, we have
  \begin{equation}
  \left(R_\phi \cdot\left(\mathcal{F} v_m\right)\right)_{k, l}=\sum_{j=1}^{d_v} R_{\phi k, l, j}\left(\mathcal{F} v_m\right)_{k, j},\quad k=1, \ldots, k_{\max }, \quad j=1, \ldots, d_v.
  \label{eq11}
  \end{equation}

\subsection{The implicit U-Net enhanced Fourier neural operator (IUFNO)}

The U-Net structure is a CNN-based network featured by the symmetrical encoder and decoder structure \cite{Ronneberger2015}. By incorporating skip connections, the U-Net enables direct transmission of feature maps from the encoder to the decoder. Since its appearance, U-Net has attracted increasing attentions due to its ability to access low-level information and high-level features simultaneously \cite{Ronneberger2015,Chen2019,Deng2023}. To better utilize the small-scale information, Li et al. has incorporated a U-Net structure to the FNO framework \cite{Li2023}. Meanwhile, the consecutive Fourier layers have also been replaced by an implicit looping Fourier layer, leading to the implicit U-Net enhanced FNO (IUFNO). Consequently, the number of trainable parameters and memory cost can be effectively reduced while the accuracy is still maintained \cite{Li2023}. The architecture of IUFNO is shown in Fig.~\ref{fig_config}b. The lifting layer $P$ and final projecting layers $Q$ are the same as those in the FNO. The iterative updating procedure of $v(\mathbf{x})$ in the IUFNO framework can be written as
\begin{equation}
  \begin{aligned}
    {v}(\mathbf{x},(l+1) dT) & =\mathcal{L}^{IUFNO}({v}(\mathbf{x}, l dT)):={v}(\mathbf{x}, l dT)+dT \sigma\left(c(\mathbf{x}, l dT) \right),\quad \forall \mathbf{x} \in D,
  \end{aligned}
  \label{vifno}
\end{equation}
where $dT$ denotes the implicit iteration steps for the Fourier layer and $l$ stands for the $l$th iteration \cite{Li2023}, and
\begin{equation}
  c(x, l dT) := W {v}(\mathbf{x}, l dT)+ \mathcal{F}^{-1}\left(R_\phi \cdot\left(\mathcal{F} {v}(\mathbf{x}, l dT)\right)\right)(\mathbf{x})+\mathcal{U^*} s(\mathbf{x}, l dT),\quad \forall \mathbf{x} \in D,
  \label{cifno}
\end{equation}
with
\begin{equation}
  s(\mathbf{x},l dT) := v(\mathbf{x},l dT) - \mathcal{F}^{-1}\left(R_\phi \cdot\left(\mathcal{F} v(\mathbf{x},l dT)\right) \right)(\mathbf{x}),\quad \forall \mathbf{x} \in D.
  \label{sifno}
\end{equation}
As can be seen, $c(\mathbf{x},l dT)\in\mathbb{R}^{d_v}$ contains both the large-scale information of the flow field learned by the Fourier layer and the small-scale information $\mathcal{U^*} s(\mathbf{x}, l dT)$ learned by the U-Net which is denoted by $\mathcal{U}^*$. Here, $s(\mathbf{x},l dT)\in\mathbb{R}^{d_v}$ is obtained by subtracting the large-scale information from full-field information $v(\mathbf{x},l dT)$ as given by Eq.~\ref{sifno}.

\section{Numerical tests in the three-dimensional turbulent channel flows}

To predict the LES solution using FNO for wall-bounded turbulence, the coarsened flow fields of the filtered DNS (fDNS) for 3D turbulent channel flows are used for the training of the FNO and IUFNO models. In the \textit{a posteriori} tests, we evaluate the FNO- and IUFNO-based solutions using initial conditions different from those in the training set, and compare the performance against the benchmark fDNS results as well as the traditional LES models.

In the current section, the DNS database and training data set are first introduced, followed by the \emph{a posteriori} test in the LES.

\subsection{The DNS database and the configuration of the training data set}

\begin{figure}\centering
\includegraphics[height=.36\textwidth]{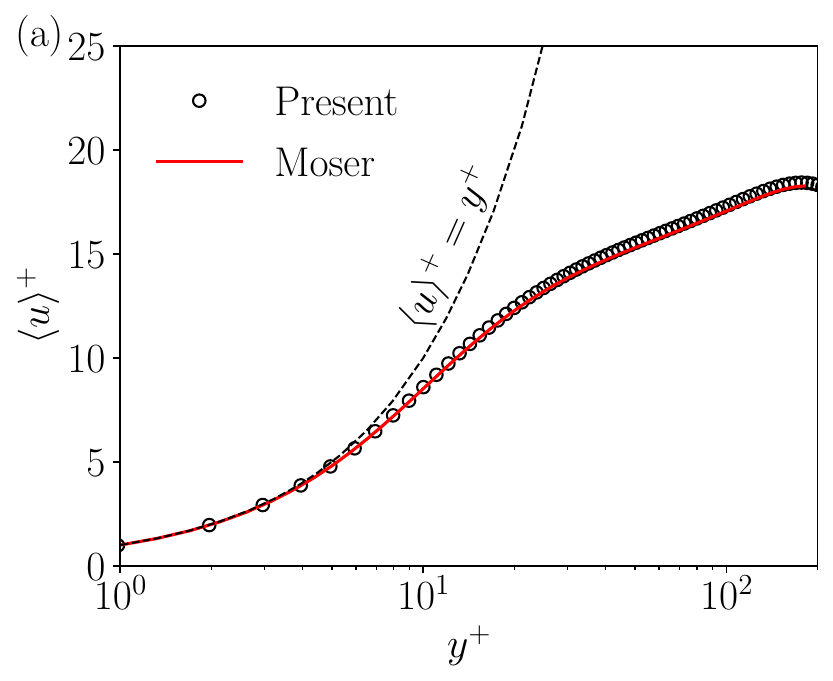}\hspace{-0.08in}
\includegraphics[height=.36\textwidth]{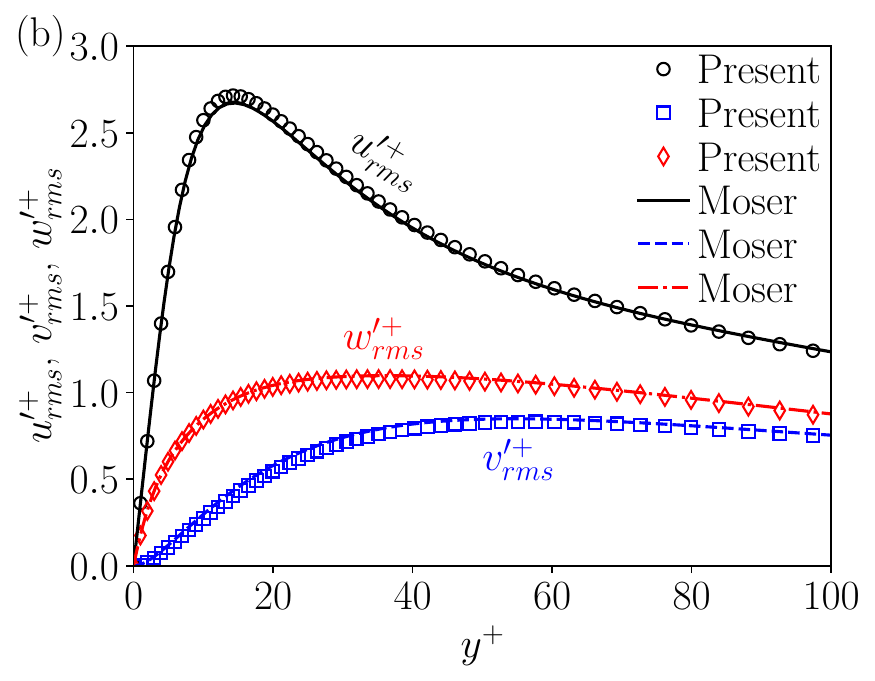}\hspace{-0.08in}

\includegraphics[height=.36\textwidth]{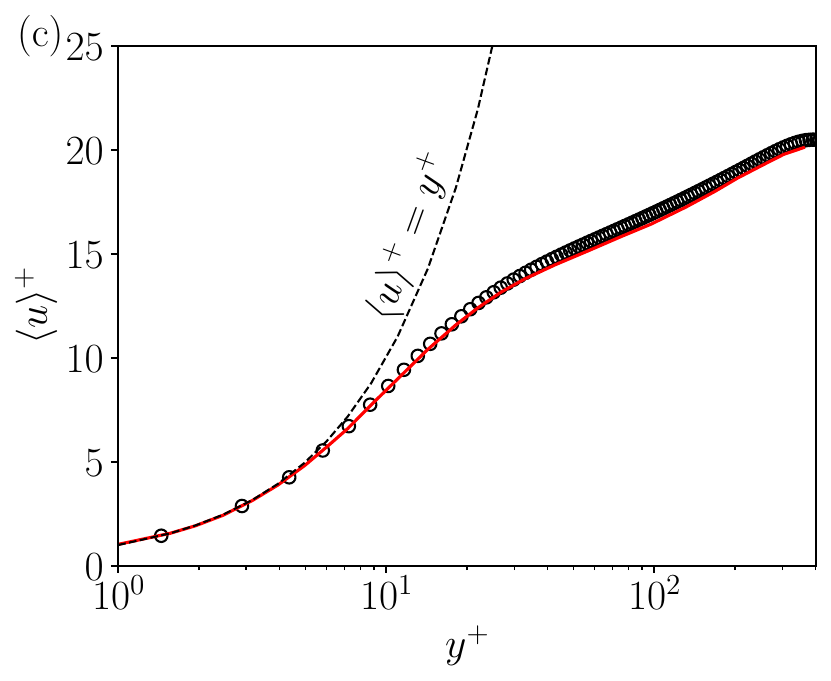}\hspace{-0.08in}
\includegraphics[height=.36\textwidth]{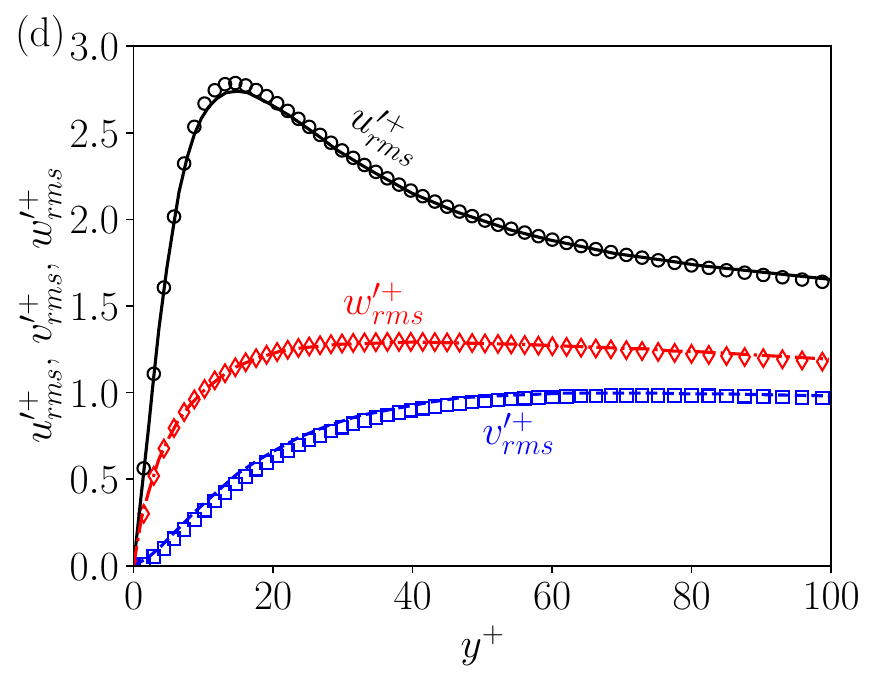}\hspace{-0.08in}

\includegraphics[height=.36\textwidth]{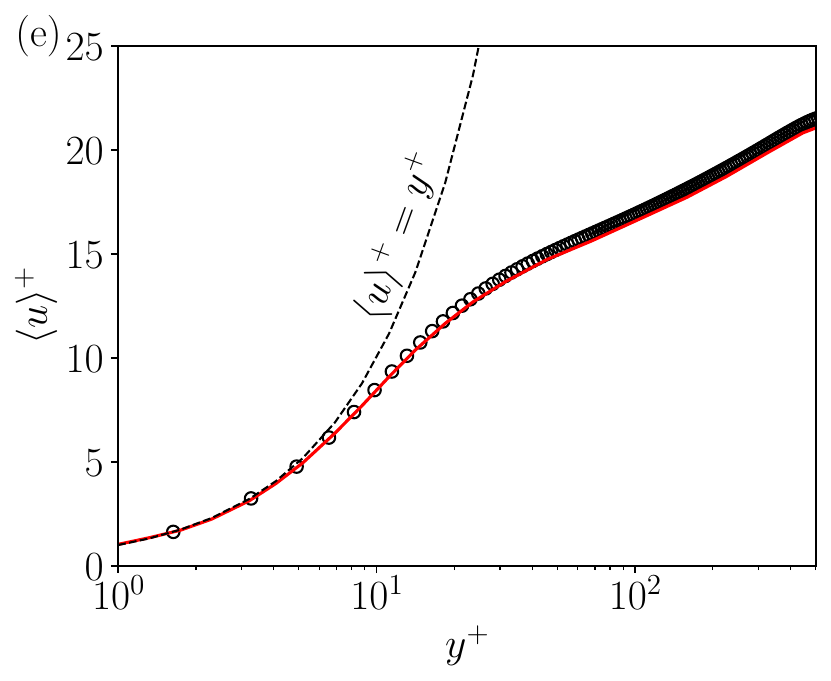}\hspace{-0.08in}
\includegraphics[height=.36\textwidth]{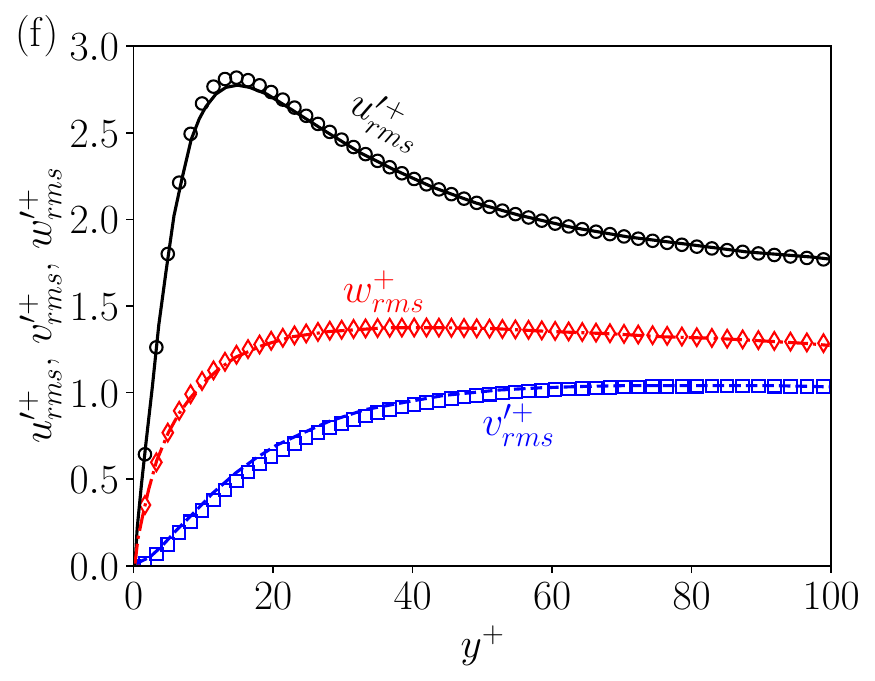}\hspace{-0.08in}
 \caption{The mean streamwise velocity and the three fluctuating velocity components for the present DNS of turbulent channel flow and the corresponding results in Ref \cite{Moser1999}: (a) mean streamwise velocity at $Re_{\tau}\approx180$; (b) velocity fluctuations at $Re_{\tau}\approx180$; (c) mean streamwise velocity at $Re_{\tau}\approx395$; (d) velocity fluctuations at $Re_{\tau}\approx395$; (e) mean streamwise velocity at $Re_{\tau}\approx590$; (f) velocity fluctuations at $Re_{\tau}\approx590$.}\label{fig_dns}

\end{figure}

The DNS database of turbulent channel flows are obtained using the open-source framework Xcompact3D, which is a high-order compact finite-difference flow solver \cite{Laizet2009,Bartholomew2020}. The details of the DNS parameters are shown in Table \ref{tab_dns}. $L_x$, $L_y$ and $L_z$ are the sizes of the 3D computational domain. $\Delta X^+$ and $\Delta Z^+$ are the normalized spacings in the streamwise and spanwise directions, respectively, and $\Delta Y^+_w$ is the distance of the first grid point off the wall. Here the superscript ``+" indicates a normalized quantity in wall (viscous) units, e.g. $y^+=y/\delta_{\nu}$, $u^+=u/u_{\tau}$ where $\delta_{\nu}$ and $u_{\tau}$ are the viscous length and wall-friction velocity, respectively. To check the accuracy of the DNS results, we compare the velocity statistics with the benchmark results by Moser et al \cite{Moser1999}. As Fig. \ref{fig_dns} shows, both the mean and fluctuating velocities agree well with the reference data at all three Reynolds numbers. Meanwhile, the linear law within the viscous sub-layer ($y^+<5$) is well recovered, suggesting the near-wall behavior is adequately captured by the current DNS.

\begin{table*}
\begin{center}
\small
\begin{tabular*}{0.8\textwidth}{@{\extracolsep{\fill}}ccccccc}
\hline
Reso. &$Re_{\tau}$ &$Lx\times Ly\times Lz$ &$ \nu$ &$\Delta X^+$ &$\Delta Y^+_w$ &$\Delta Z^+$ \\ \hline
$192\times129\times64$ &180 &$4\pi\times2\times4\pi/3$ &1/4200 &11.6 &0.98 &11.6 \\ 
$256\times193\times128$ &395 &$4\pi\times2\times4\pi/3$ &1/10500 &19.1 &1.4 &12.8 \\ 
$384\times257\times192$ &590 &$4\pi\times2\times4\pi/3$ &1/16800 &19.3 &1.6 &12.9\\ \hline
\end{tabular*}
\normalsize
\caption{Parameters for the direction numerical simulation of turbulent channel flow.}\label{tab_dns}
\end{center}
\end{table*}

As discussed, the filtered DNS (fDNS) data is the benchmark for the resolved variables in LES. To obtain the training data set, the DNS data of turbulent channel flows are filtered and then coarsened to the LES grids. In order to reduce the computational cost, the adopted LES grids are tangibly coarser than the general requirements for wall-resolved LES, which can be close to the grid requirement for DNS \cite{Yang2021}. In current work, the LES grids are $32\times33\times16$, $64\times49\times32$ and $64\times65\times32$ for $Re_{\tau}\approx180$, $395$ and $590$, respectively. The details of the LES parameters are shown in Table \ref{tab_les}. As can be seen, the grid sizes are larger compared to those for wall-resolved LES \cite{Yang2021,Xu2023}. The DNS data is filtered in the homogeneous $x$ and $z$ directions using a box filter \cite{Pope2000}, and the corresponding filter widths are equal to the LES grid sizes $\Delta_x$ and $\Delta_z$, respectively. In the training data set, 400 snapshots of the fDNS data are extracted at every $\Delta T=200 \Delta t$, where the DNS time step $\Delta t=0.005$. Hence, $\Delta T=1$ in non-dimensional unit. Defining the wall viscous time unit as $\tau_v=\delta_v/u_{\tau}$, we have $\Delta T=7.5\tau_v$, $14.6\tau_v$ and $20.7\tau_v$ for $Re_{\tau}\approx180$, $395$ and $590$, respectively. Based on our test, such configuration can give the best long-term performance. Meanwhile, 20 groups of fDNS data sets are generated using different initial conditions to enlarge the training database.

\begin{table*}
\begin{center}
\small
\begin{tabular*}{0.8\textwidth}{@{\extracolsep{\fill}}ccccccc}
\hline
Reso. &$Re_{\tau}$ &$Lx\times Ly\times Lz$ &$ \nu$ &$\Delta X^+$ &$\Delta Y^+_w$ &$\Delta Z^+$ \\ \hline
$32\times33\times16$ &180 &$4\pi\times2\times4\pi/3$ &1/4200 &69.6 &3.93 &46.4 \\ 
$64\times49\times32$ &395 &$4\pi\times2\times4\pi/3$ &1/10500 &76.4 &5.6 &51.2 \\ 
$64\times65\times32$ &590 &$4\pi\times2\times4\pi/3$ &1/16800 &115.8 &6.4 &77.4\\ \hline
\end{tabular*}
\normalsize
\caption{Parameters for the large-eddy simulation of turbulent channel flow.}\label{tab_les}
\end{center}
\end{table*}

It is important to note that the FNO can accommodate non-periodic boundary conditions in the $y$ normal direction of channel flow, thanks to the bias term $W$ (cf. Fig. \ref{fig_config}). However, to further alleviate the effects of the non-periodicity, the mean velocity field of fDNS is subtracted from the training data such that only the fluctuations are used for training. Meanwhile, to apply the fast Fourier transform (FFT), the original coordinates in the wall-normal y direction should be remapped onto uniform coordinates \cite{Li2022a,Deng2023,Meng2023}. For channel flow, while the mesh is nonuniform in $y$ direction, it is still a structured mesh and can be transformed into a uniform mesh by the mapping $[y_1,y_2,y_3,\ldots]\longmapsto[y_1/r_1,y_2/r_2,y_3/r_3,\ldots]$ \cite{Li2022a}. Hence the FFT can still be conveniently applied.

For both the FNO and IUFNO, the known velocity fields of several time nodes are taken as the input to the model. Meanwhile, the increment of the fluctuating field is learned \cite{Li2022,Li2023}. By letting $U_m$ denote the $m$th time-node velocity, the $m$th increment (i.e. the difference of velocity field between two adjacent time nodes) can be written as $\Delta U_m=U_{m+1}-U_m$. The fDNS data of the previous five time nodes $[U_{m-4}, U_{m-3}, U_{m-2}, U_{m-1}, U_m]$ are taken as the model input and $\Delta U_m$ is taken as the output. Then $U_{m+1}$ can be obtained by $U_{m+1}=U_{m}+\Delta U_m$. In this case, 400 time-nodes in each of the 20 groups can generate 395 input-output pairs. 80\% of the data are used for training and the rest for on-line testing.

\begin{figure}\centering
\includegraphics[height=.36\textwidth]{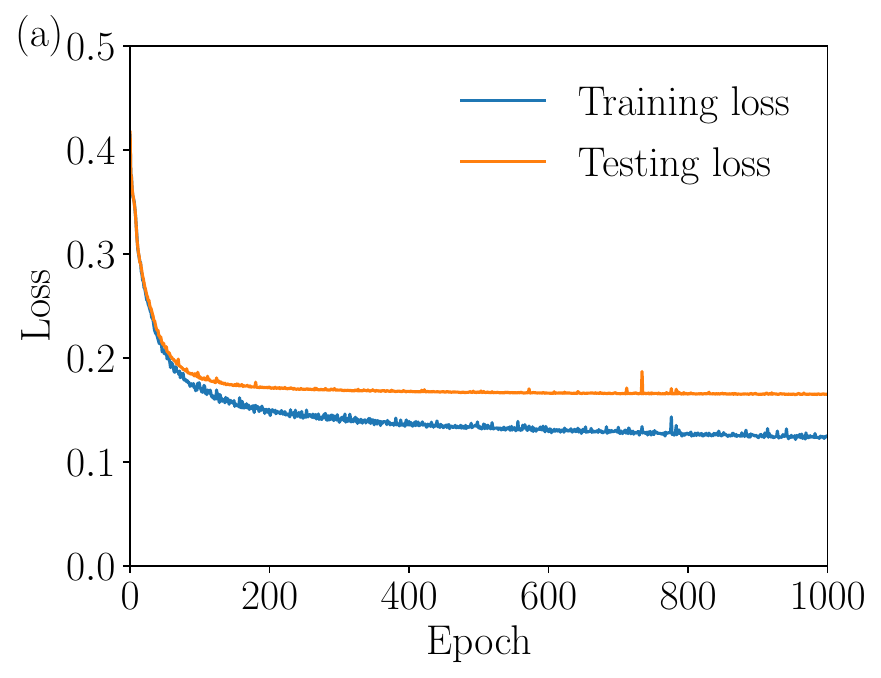}\hspace{-0.08in}
\includegraphics[height=.36\textwidth]{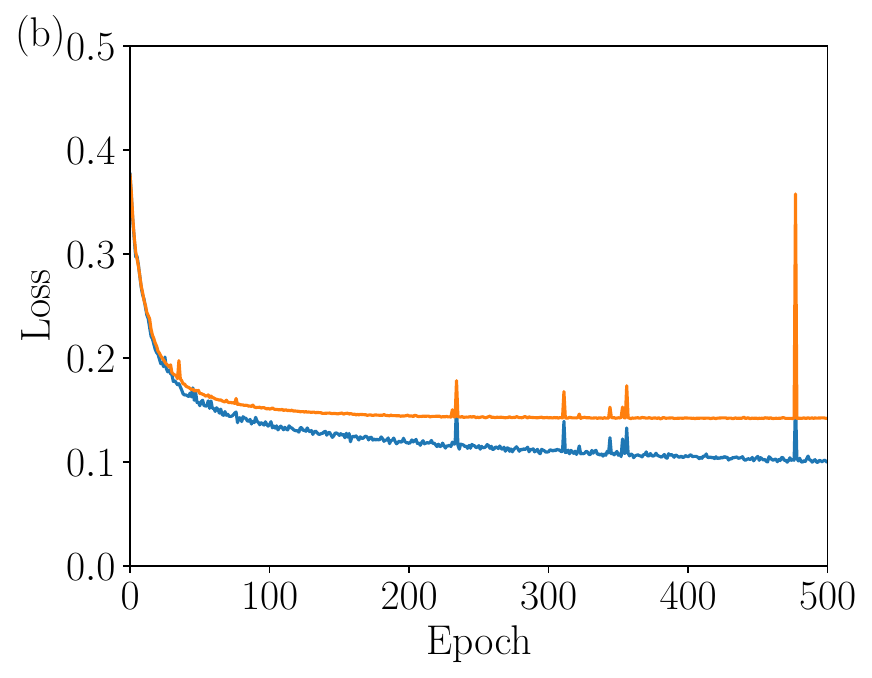}\hspace{-0.08in}
\includegraphics[height=.36\textwidth]{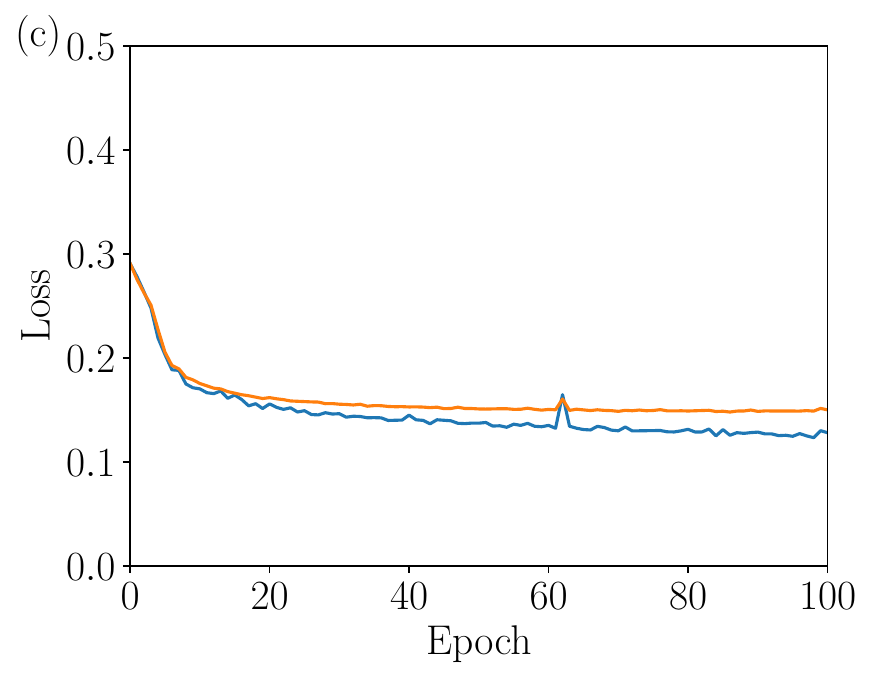}\hspace{-0.08in}
\includegraphics[height=.36\textwidth]{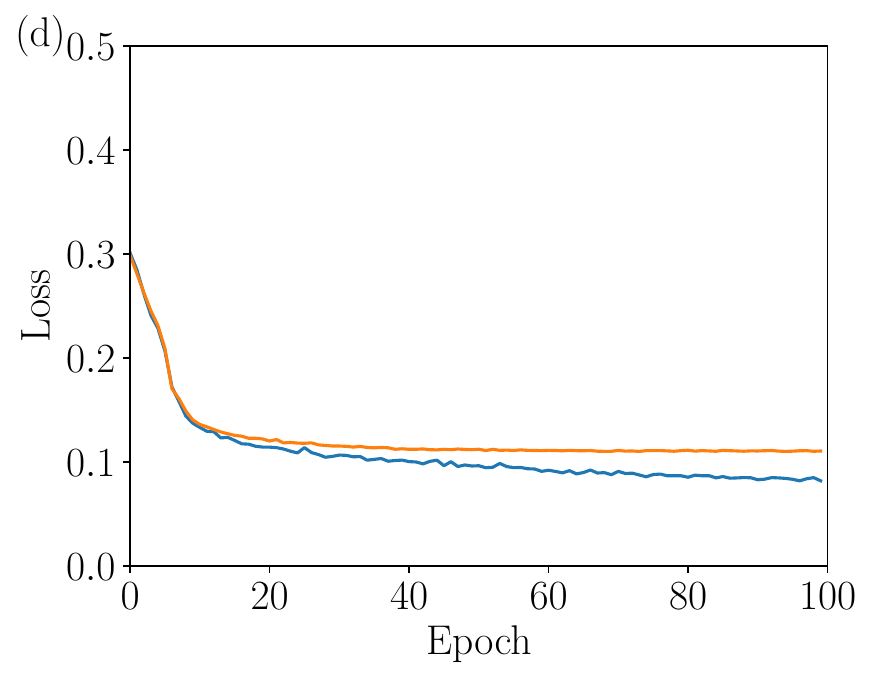}\hspace{-0.08in}
 \caption{The evolutions of the loss curves: (a) FNO; (b) IFNO; (c) UFNO; (d) IUFNO.}\label{fig_loss}
\end{figure}

The training setups are configured as follows: For both the FNO and IUFNO models, two fully-connected layers (i.e. $P$ and $Q$ in Fig. \ref{fig_config}) are configured before and after the Fourier layer(s). For the FNO model, four consecutive Fourier layers are adopted \cite{Li2022}. For the IUFNO model, the consecutive Fourier layers are replaced by an implicit Fourier layer with embedded U-Net structure, and the numbers of internal iterations for the Fourier layer in the IUFNO model are 40, 20 and 20 for $Re_{\tau}\approx180$, $395$ and $590$, respectively. In the U-Net structure of the IUFNO model, three consecutive encoders are configured, followed by three decoders with skip connections \cite{Ronneberger2015}. For all the trainings, the cutoff wavenumbers for the Fourier modes truncation are based on the half of the minimum grid numbers (i.e. in z direction). Hence, we set $k_{max}=8$, $16$ and $16$ for $Re_{\tau}\approx180$, $395$ and $590$, respectively. The Adam optimizer is used for optimization \cite{Kingma2014}, the initial learning rate is set to $10^{-3}$, and the GELU function is chosen as the activation function \cite{Hendrycks2016}. The training and testing losses are defined as
\begin{equation}
  \textit{Loss}=\frac{\|\Delta u^*-\Delta u\|_2}{\|\Delta u\|_2}, \text { where }\|\mathbf{A}\|_2=\frac{1}{n} \sqrt{\sum_{k=1}^n\left|\mathbf{A}_{\mathbf{k}}\right|^2}.
  \label{eqloss}
\end{equation}
Here, $\Delta u^*$ denotes the prediction of velocity increments and $\Delta u$ is the ground truth. The evolution curves of the training and testing losses at $Re_{\tau}\approx180$ are shown in Fig. \ref{fig_loss}. Here, we also present the corresponding results for the implicit FNO (IFNO) \cite{You2022} which does not incorporate the U-Net structure, and the results for the U-Net enhanced FNO (UFNO) \cite{Wen2022} which uses consecutive U-Net embedded Fourier layers instead of the implicit looping structure. As can be seen, the IUFNO model has the lowest loss compared to the other FNO models. We also observe that the U-Net-based models (UFNO and IUFNO) converge faster compared to the other models.

\subsection{The \emph{a posteriori} tests in the LES.}

\begin{figure}\centering
\includegraphics[height=.36\textwidth]{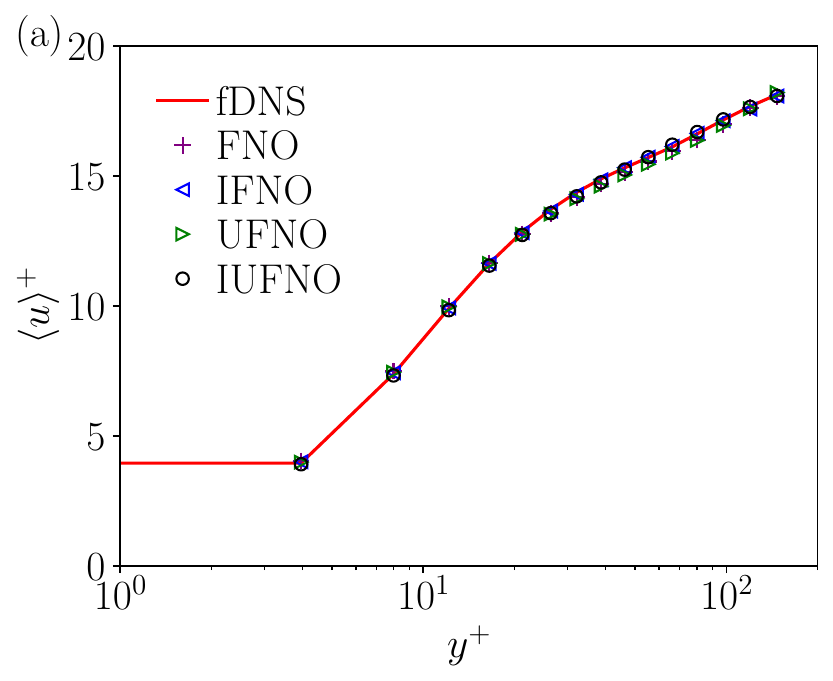}\hspace{-0.08in}
\includegraphics[height=.36\textwidth]{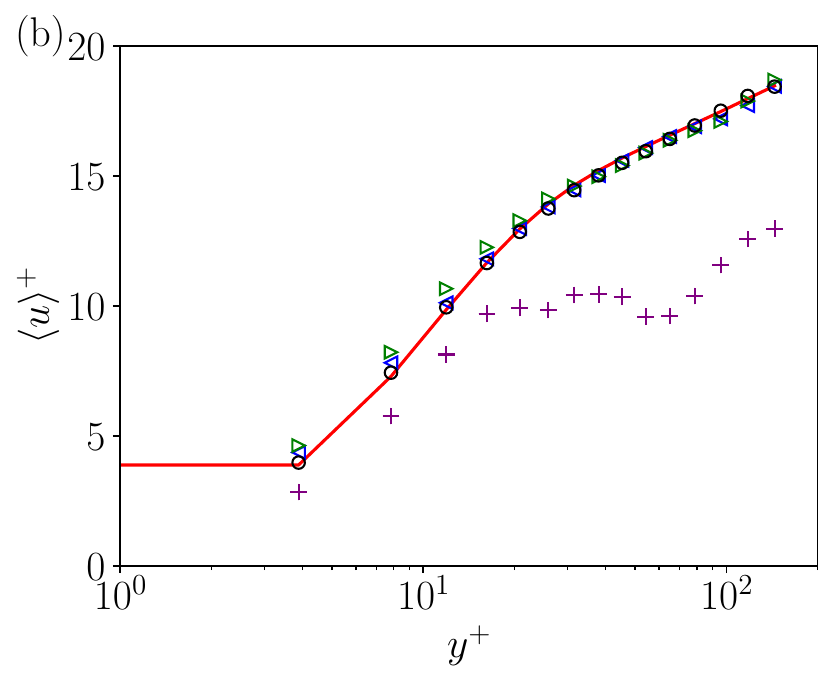}\hspace{-0.08in}

\includegraphics[height=.36\textwidth]{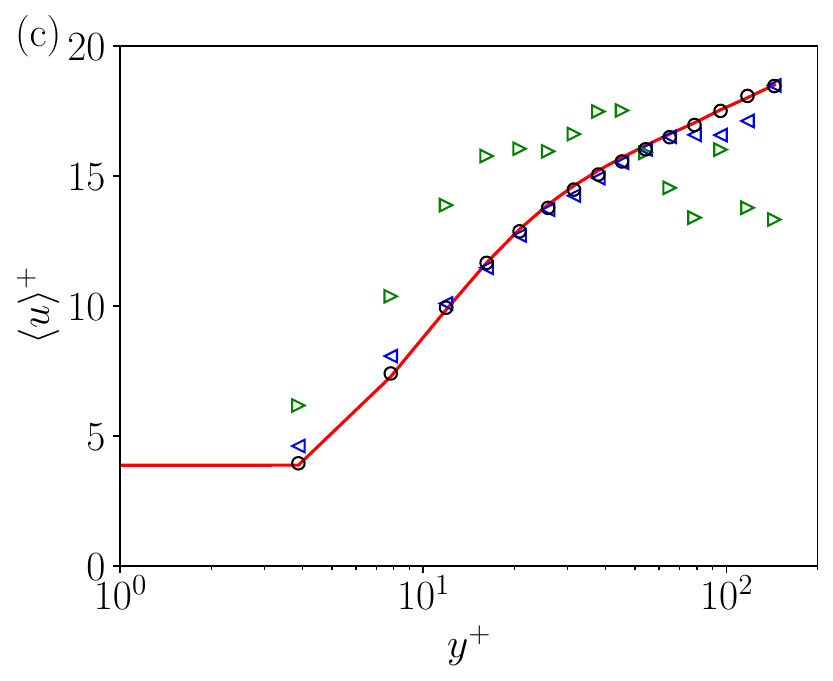}\hspace{-0.08in}
\includegraphics[height=.36\textwidth]{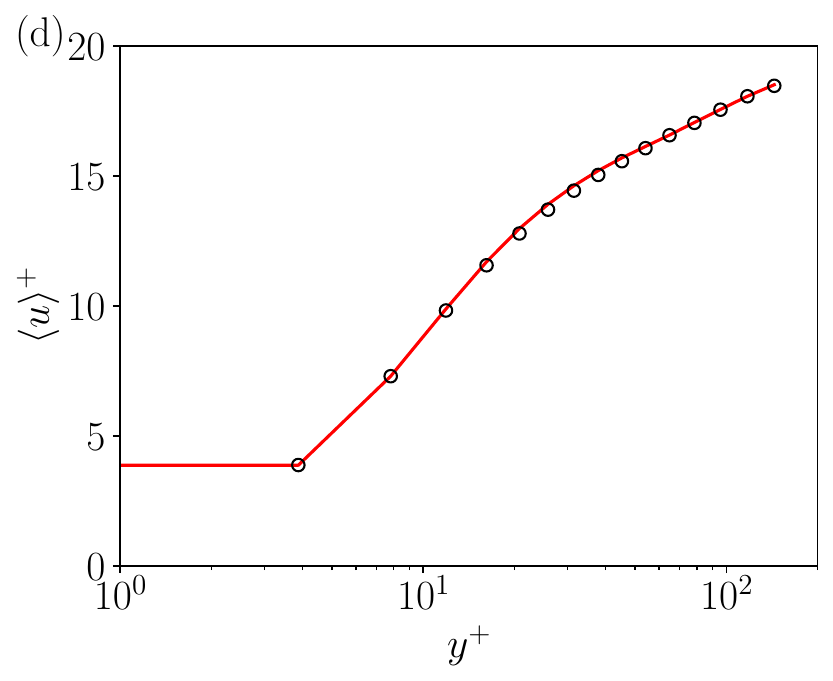}\hspace{-0.08in}

\includegraphics[height=.36\textwidth]{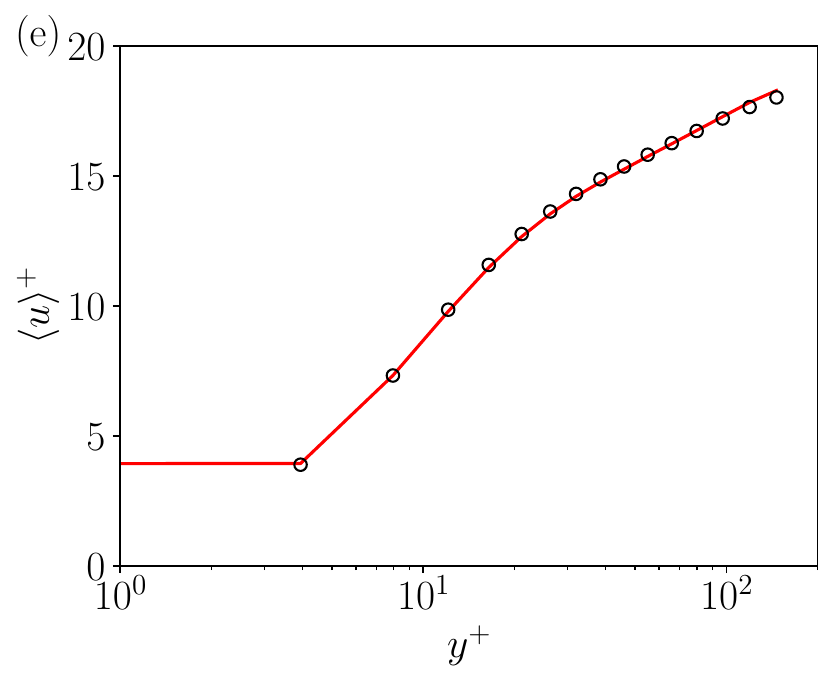}\hspace{-0.08in}
\includegraphics[height=.36\textwidth]{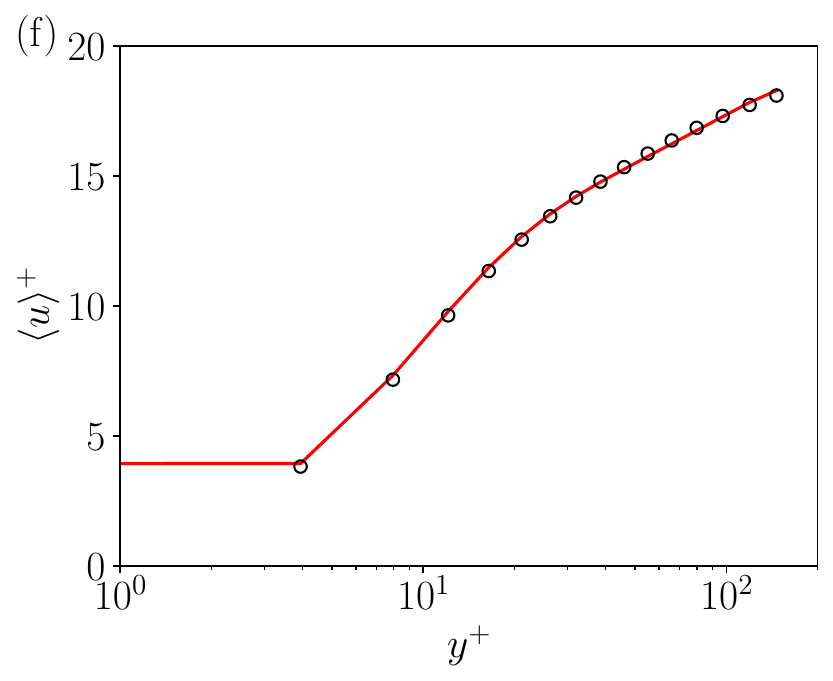}\hspace{-0.08in}

 \caption{The mean streamwise velocities at $Re_{\tau}\approx180$ for different FNO-based LES models: (a) $t=10\Delta T$; (b) $t=25\Delta T$; (c) $t=29\Delta T$; (d) $t=40\Delta T$; (e) $t=600\Delta T$; (f) $t=800\Delta T$.}\label{fig_fno_iufno}
\end{figure}

In the \emph{a posteriori} tests, the initial turbulent fields are taken from a new fDNS field which is different from those in the training set, such that we can test whether the trained model can be generalize to different initial conditions. Meanwhile, the no-slip condition is reinforced in FNO and IUFNO in the \emph{a posteriori} tests. To compare the performances of different FNO-based frameworks, we show the predicted mean streamwise velocity profiles at different time instants for $Re_{\tau}\approx180$ in Fig. \ref{fig_fno_iufno}. It can be seen that all the FNO models predict the velocity profile well initially (at $t=10\Delta T$). However, as the solutions evolve, the predictions of the FNO, IFNO and UFNO models diverge from the fDNS benchmark sooner or later, while the IUFNO model remains adequately accurate. At $t=25\Delta T$, the FNO result starts to strongly deviate from the fDNS benchmark, while the other models are still reasonably accurate. At $t=29\Delta T$, FNO diverges, and UFNO strongly deviates from the fDNS result, while IFNO only slightly deviates. At $t=40\Delta T$, all models diverge except the IUFNO model. Most importantly, at $t=600\Delta T$ and $800\Delta T$, both of which are far beyond the time range of the training set ($t<400\Delta T$), the IUFNO model remains numerically stable and accurate. In physical sense, $800\Delta T$ allows the bulk flow to pass through the channel for approximately 42.5 times, demonstrating the long-term predictive ability of the IUFNO model. Hence, in the rest of the work, only the results of IUFNO will be presented and compared against the traditional LES models including the DSM and WALE models.

\begin{figure}\centering
\includegraphics[height=.38\textwidth]{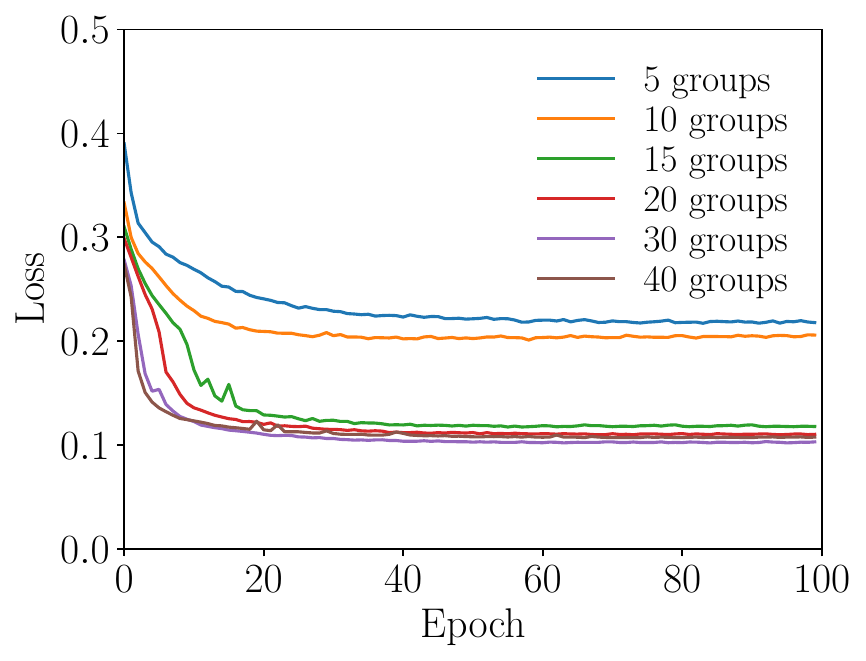}
 \caption{The evolutions of the testing loss with different amount of training data.}\label{fig_loss_converge}
\end{figure}

Since the IUFNO framework is trained on DNS data, it is interesting to study how the model converges with different amount of training data. Fig. \ref{fig_loss_converge} displays the evolutions of the testing loss with different amount of training data at $Re_{\tau}=180$. As can be seen, when the amount of training data is small (5 and 10 groups of DNS), the testing losses only decrease to the level of 0.2 to 0.25. As the training data set increases to 15 groups, a significant reduction of the testing loss can be observed as it decreases from 0.2 to about 0.1. Further enlargement of the training data set does not alter much the testing loss, which remains at the level of 0.1. Hence, 20 groups of the training DNS data are adopted in the current work.

\begin{figure}\centering
\includegraphics[height=.36\textwidth]{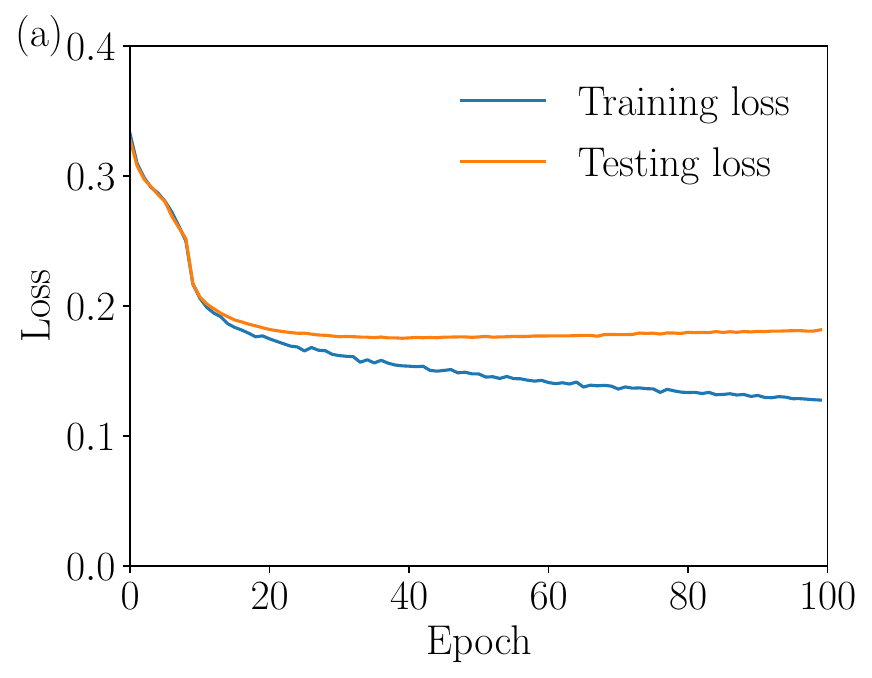}\hspace{-0.10in}
\includegraphics[height=.36\textwidth]{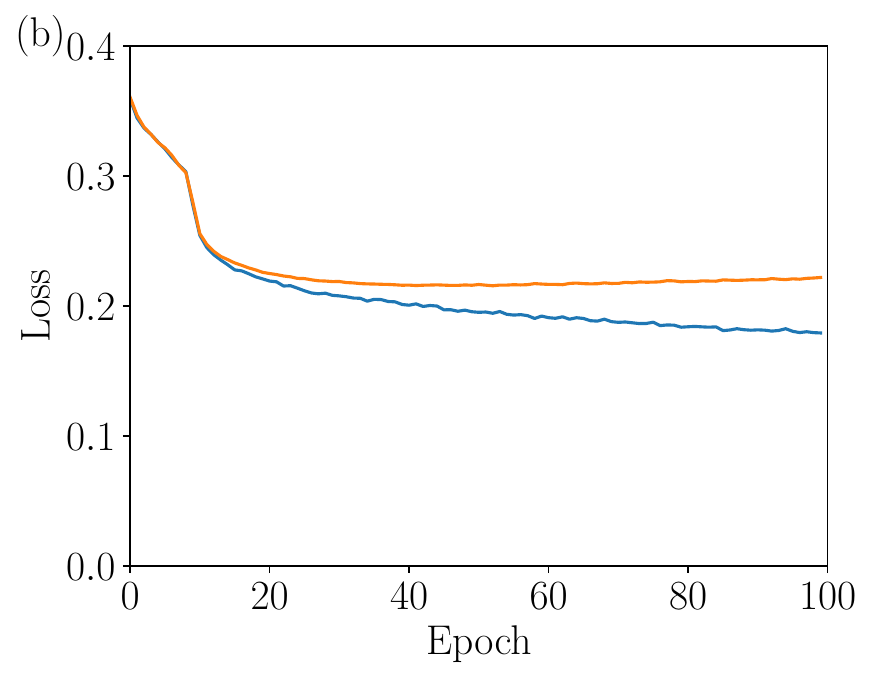}\hspace{-0.10in}
 \caption{The evolutions of the loss curves for IUFNO: (a) $Re_{\tau}\approx395$; (b) $Re_{\tau}\approx590$.}\label{fig_loss_395_590}
\end{figure}

The training and testing losses for $Re_{\tau}=395$ and $590$ are shown in Figs. \ref{fig_loss_395_590}a and \ref{fig_loss_395_590}b, respectively. It can be seen that as $Re_{\tau}$ increases, the loss goes up, indicating the increased difficulty in learning high Reynolds number turbulent flow. Meanwhile, to avoid over-fitting of the model, we have applied early stopping which is widely used in gradient descent learning \cite{Yao2007,Raskutti2011}. In both cases, the model parameters are extracted after 30 epochs before the performance on the testing set starts to degrade. 

\begin{figure}\centering
\includegraphics[height=.36\textwidth]{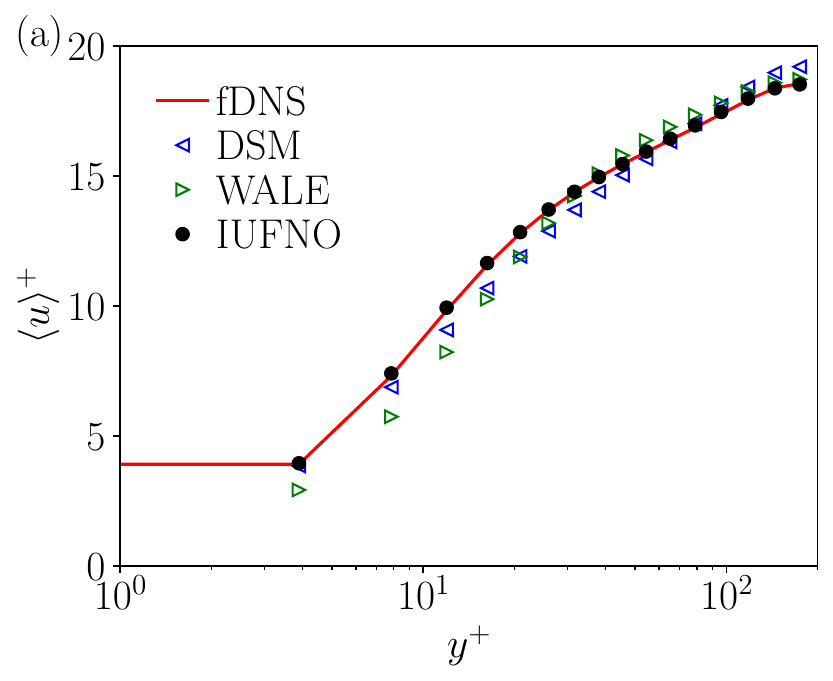}\hspace{-0.00in}
\includegraphics[height=.36\textwidth]{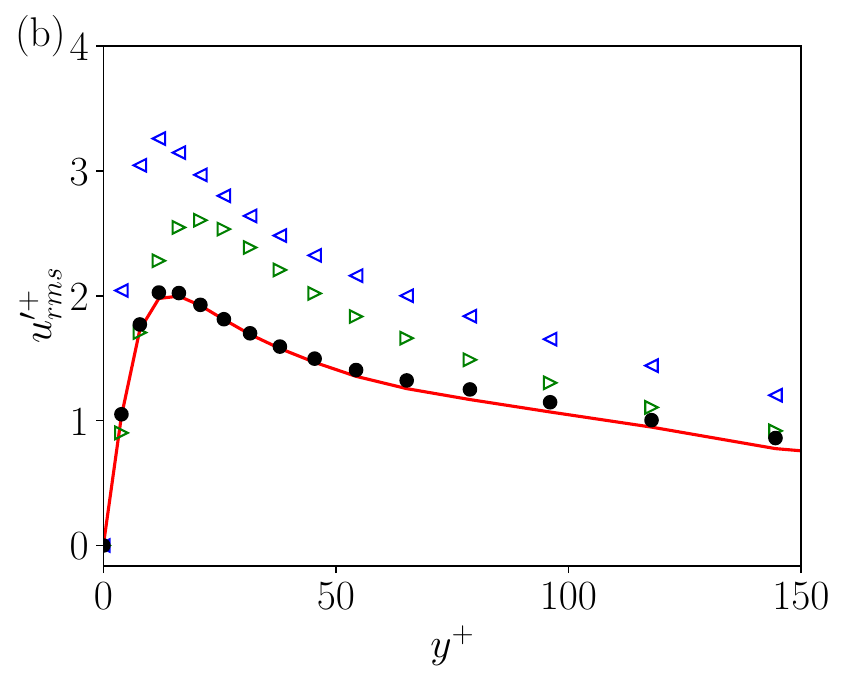}\hspace{-0.05in}

\includegraphics[height=.36\textwidth]{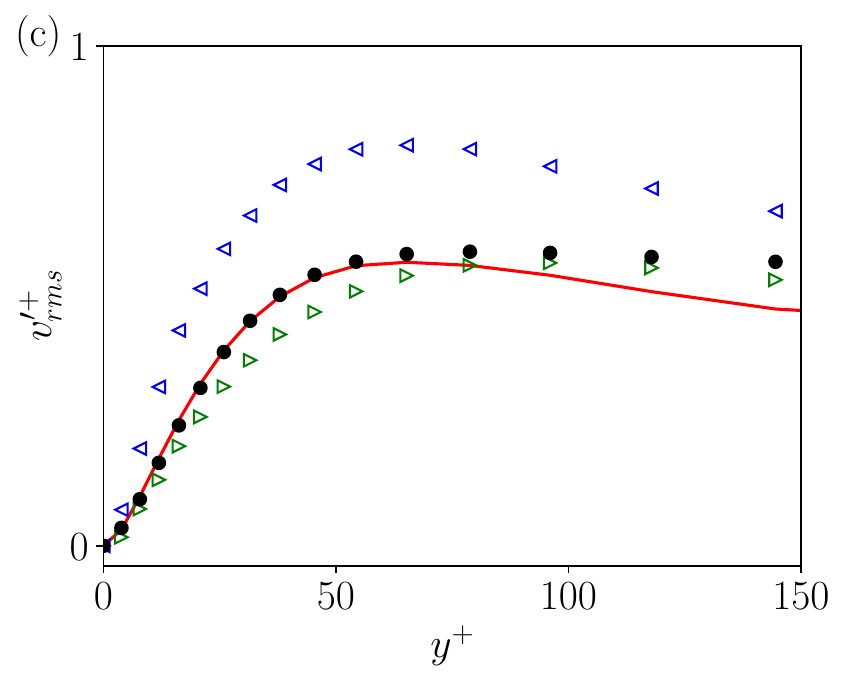}\hspace{-0.10in}
\includegraphics[height=.36\textwidth]{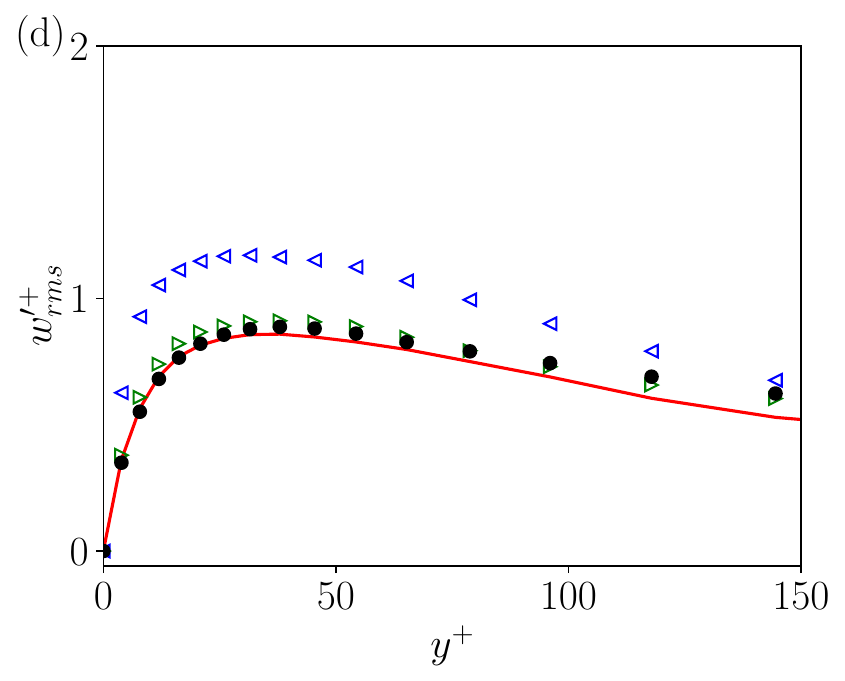}\hspace{-0.10in}

 \caption{The mean streamwise velocity and rms fluctuating velocities at $Re_{\tau}\approx180$: (a) mean streamwise velocity; (b) rms fluctuation of streamwise velocity; (c) rms fluctuation of transverse velocity; (d) rms fluctuation of spanwise velocity.}\label{fig_vel_180}
\end{figure}

\begin{figure}\centering
\includegraphics[height=.36\textwidth]{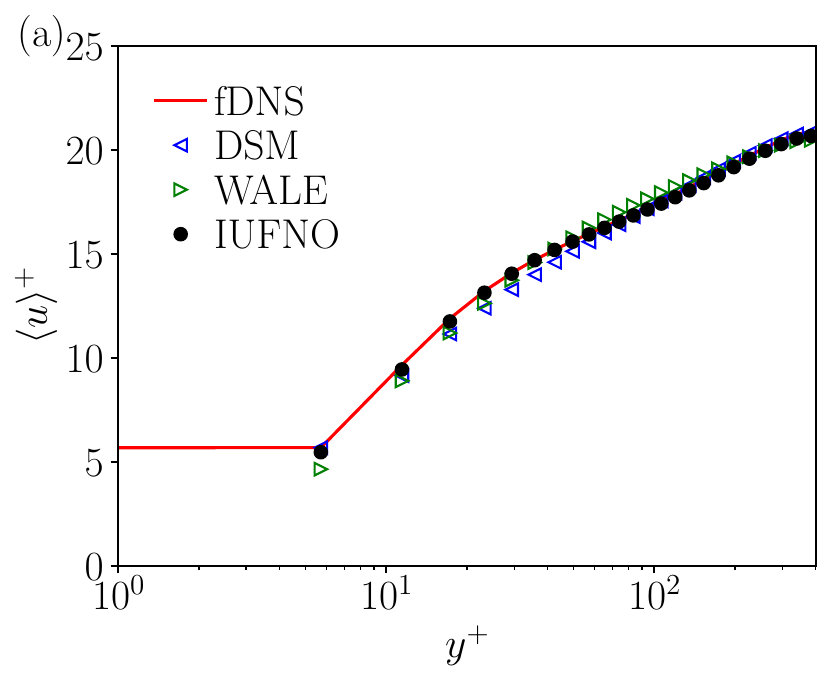}\hspace{-0.00in}
\includegraphics[height=.36\textwidth]{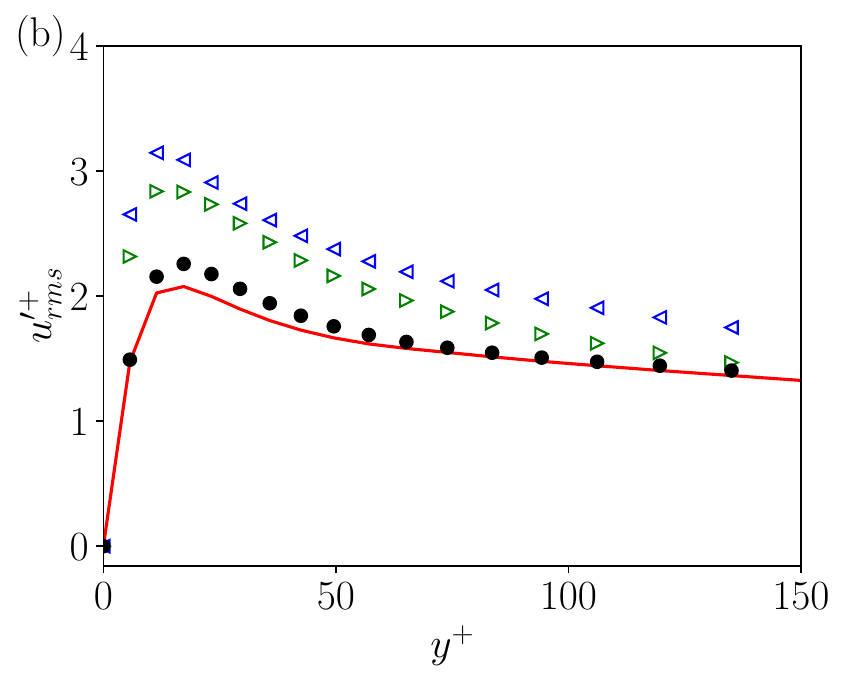}\hspace{-0.05in}

\includegraphics[height=.36\textwidth]{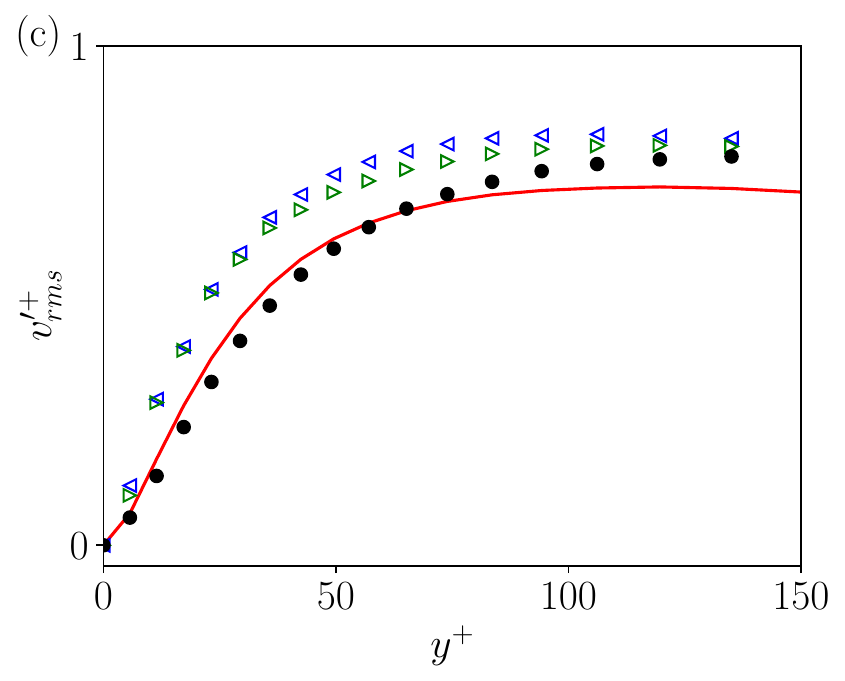}\hspace{-0.10in}
\includegraphics[height=.36\textwidth]{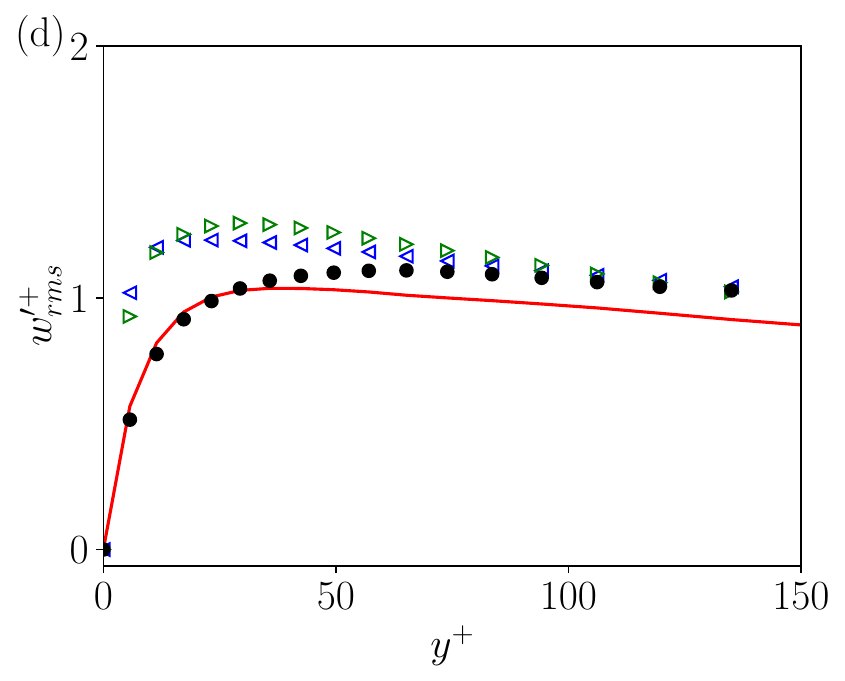}\hspace{-0.10in}

 \caption{The mean streamwise velocity and rms fluctuating velocities at $Re_{\tau}\approx395$: (a) mean streamwise velocity; (b) rms fluctuation of streamwise velocity; (c) rms fluctuation of transverse velocity; (d) rms fluctuation of spanwise velocity.}\label{fig_vel_395}
\end{figure}

\begin{figure}\centering
\includegraphics[height=.36\textwidth]{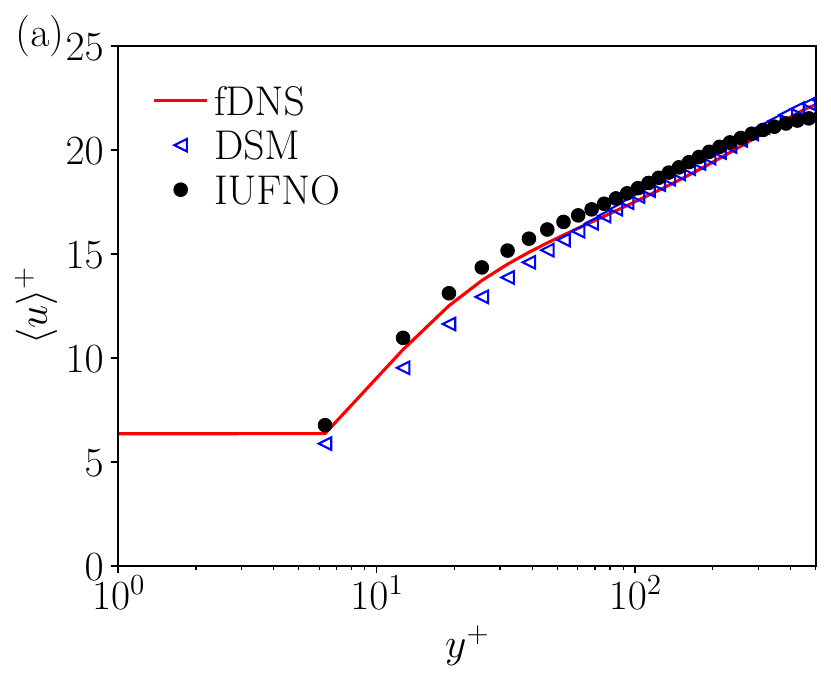}\hspace{-0.00in}
\includegraphics[height=.36\textwidth]{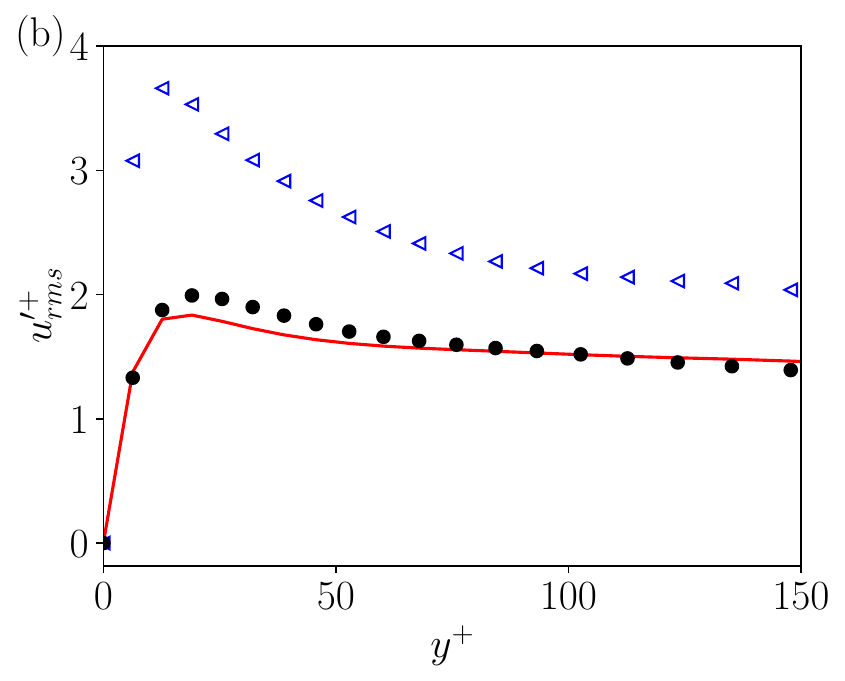}\hspace{-0.05in}

\includegraphics[height=.36\textwidth]{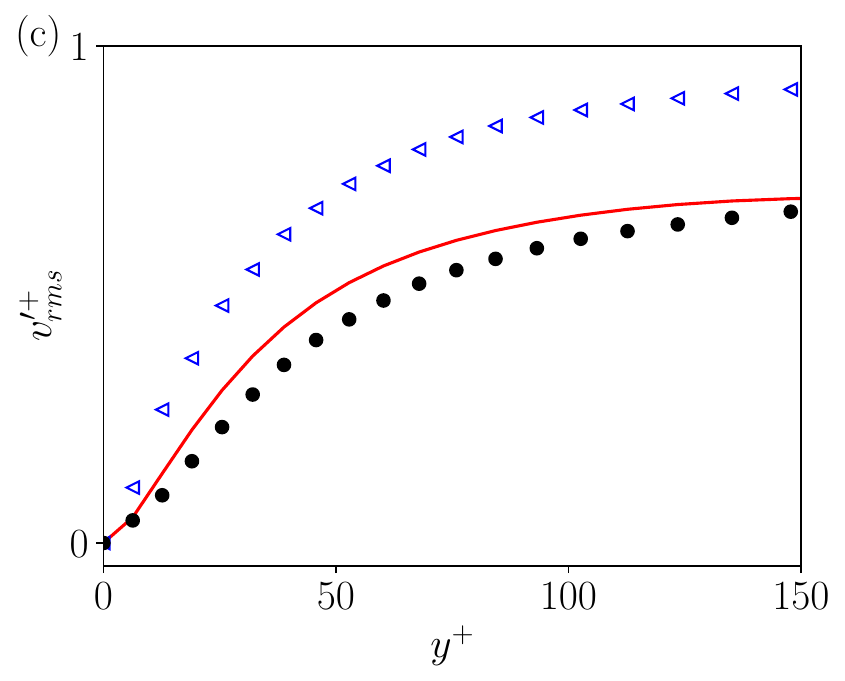}\hspace{-0.10in}
\includegraphics[height=.36\textwidth]{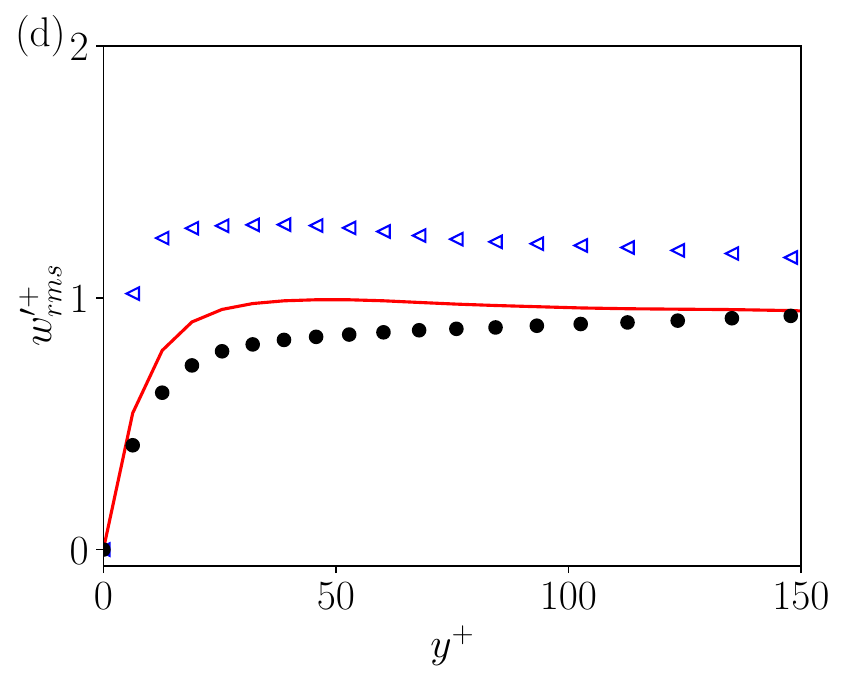}\hspace{-0.10in}

 \caption{The mean streamwise velocity and rms fluctuating velocities at $Re_{\tau}\approx590$: (a) mean streamwise velocity; (b) rms fluctuation of streamwise velocity; (c) rms fluctuation of transverse velocity; (d) rms fluctuation of spanwise velocity.}\label{fig_vel_590}
\end{figure}

In Figs. \ref{fig_vel_180}, \ref{fig_vel_395} and \ref{fig_vel_590}, the mean streamwise velocity and root-mean-squared (rms) fluctuating velocities predicted by the IUFNO model are displayed for $Re_{\tau}\approx180$, $395$ and $590$, respectively. Also shown in the figures are the corresponding predictions by the DSM and WALE models. As can be seen, the IUFNO model performs reasonably better at all three Reynolds numbers compared to the DSM and WALE models. Nevertheless, the prediction of the mean velocity by IUFNO at $Re_{\tau}\approx590$ is not as good as that at $Re_{\tau}\approx180$ and $395$. This also reflects the increasing difficulty in predicting turbulent flows at high Reynolds number. For the traditional LES models, we observe that the WALE model performs tangibly better than the DSM. Meanwhile, the WALE model diverges at $Re_{\tau}\approx590$, thus the results of WALE are not presented for $Re_{\tau}\approx590$. Here, it should be emphasized that the current LES grids are coarser than the general requirements for wall-resolved LES \cite{Yang2021,Xu2023}. In the expense of additional computational cost (i.e. using sufficiently fine grids or carefully configured wall model), the traditional LES models can also achieve reasonably accurate LES predictions, which is not the concern of the current work.

\begin{figure}\centering
\includegraphics[height=.33\textwidth]{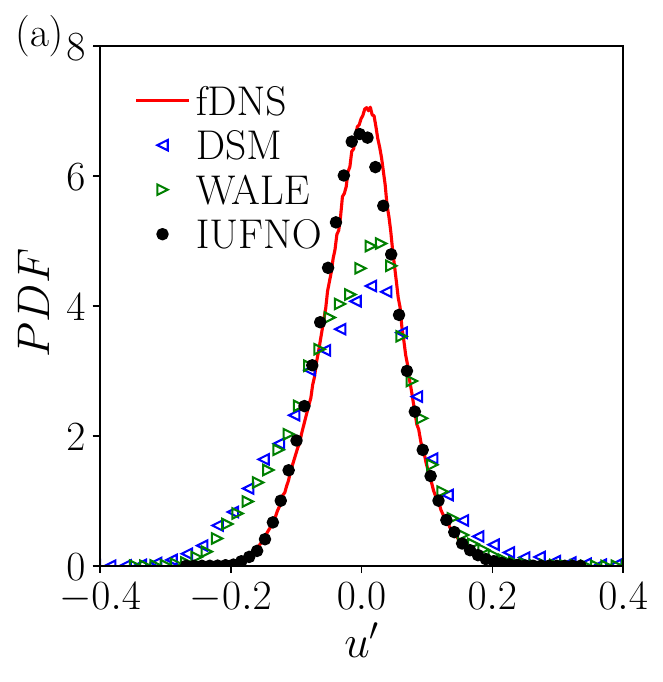}\hspace{-0.10in}
\includegraphics[height=.33\textwidth]{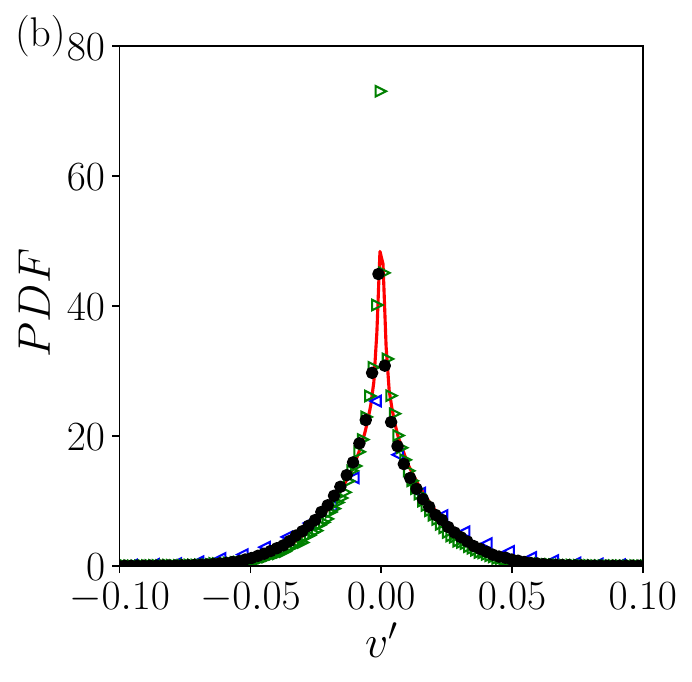}\hspace{-0.10in}
\includegraphics[height=.33\textwidth]{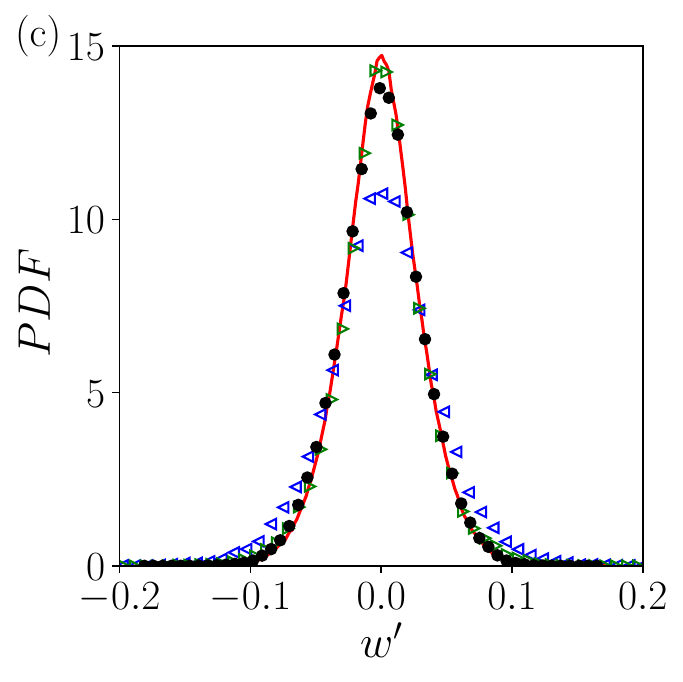}\hspace{-0.10in}

\includegraphics[height=.33\textwidth]{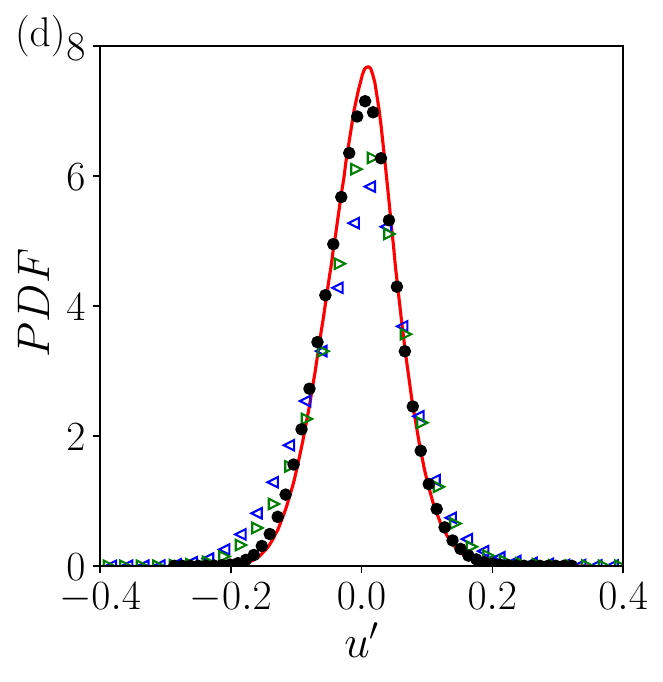}\hspace{-0.10in}
\includegraphics[height=.33\textwidth]{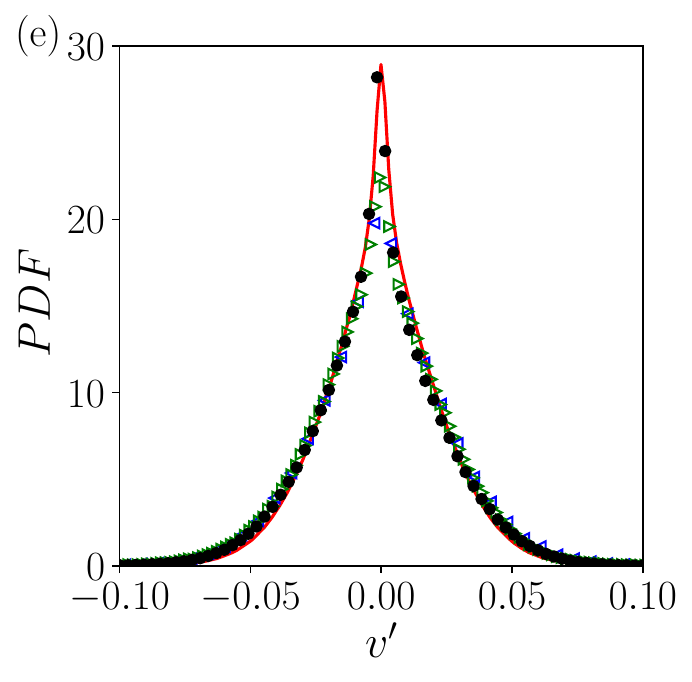}\hspace{-0.10in}
\includegraphics[height=.33\textwidth]{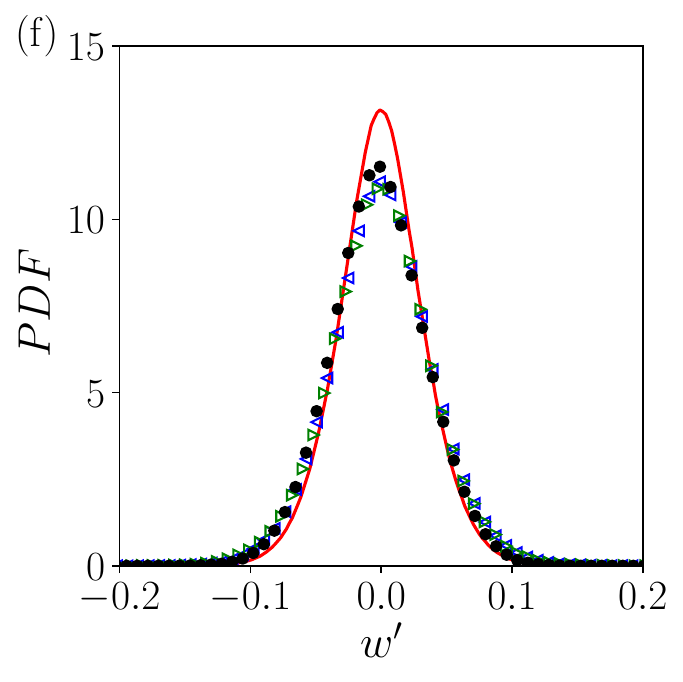}\hspace{-0.10in}

\includegraphics[height=.33\textwidth]{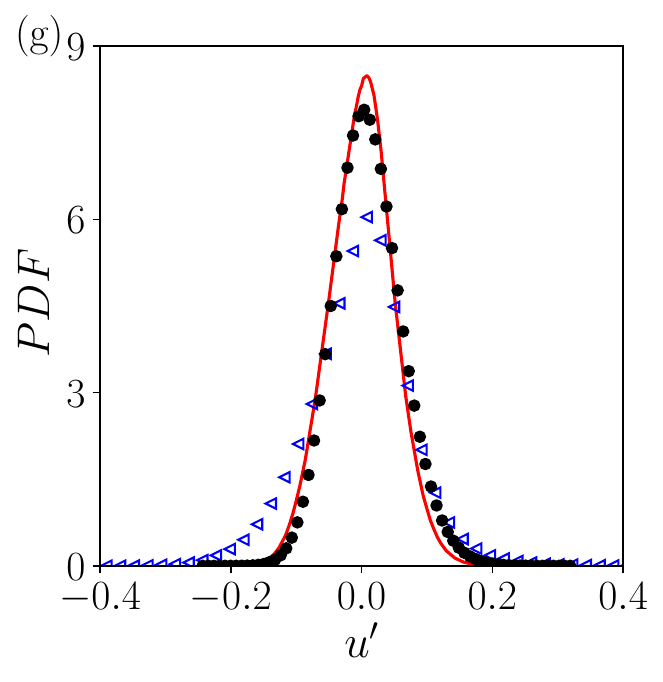}\hspace{-0.10in}
\includegraphics[height=.33\textwidth]{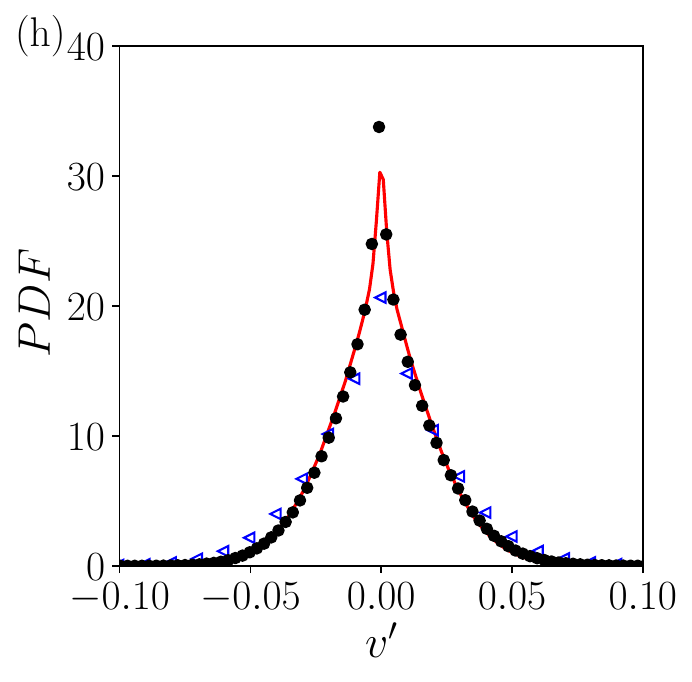}\hspace{-0.10in}
\includegraphics[height=.33\textwidth]{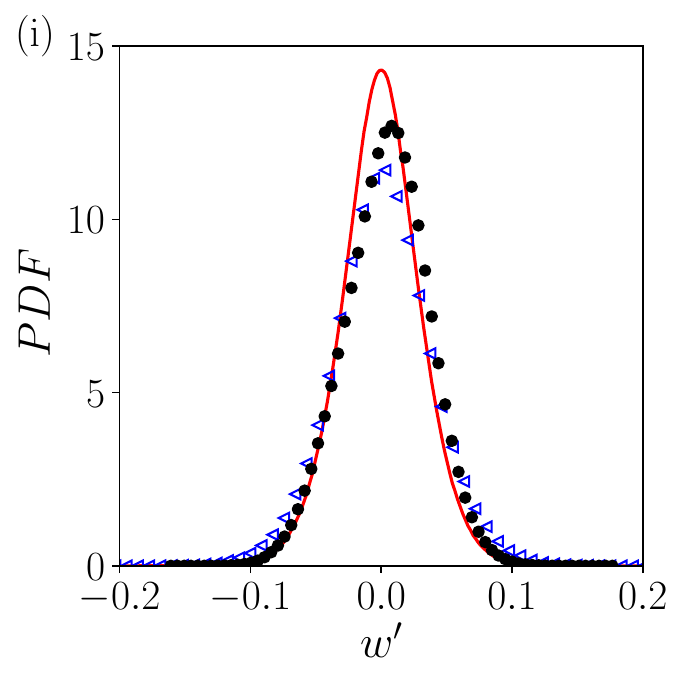}\hspace{-0.10in}

\caption{The probability density functions (PDFs) of the velocity fluctuation components at different Reynolds numbers: (a) $u', Re_{\tau}\approx180$; (b) $v', Re_{\tau}\approx180$; (c) $w', Re_{\tau}\approx180$; (d) $u', Re_{\tau}\approx395$; (e) $v', Re_{\tau}\approx395$; (f) $w', Re_{\tau}\approx395$; (g) $u', Re_{\tau}\approx590$; (h) $v', Re_{\tau}\approx590$; (i) $w', Re_{\tau}\approx590$.}\label{fig_pdfvel}
\end{figure}

The probability density functions (PDFs) of the three fluctuating velocity components are shown in Fig. \ref{fig_pdfvel} for different friction Reynolds numbers. The PDFs of the velocity fluctuations are all symmetric, and the ranges of the PDFs are in consistency with the corresponding rms values in Figs. \ref{fig_vel_180}, \ref{fig_vel_395} and \ref{fig_vel_590}. It can be seen that the IUFNO model gives reasonably good predictions of the PDFs at all three Reynolds numbers. Notably, the advantage of the IUFNO model is more pronounced for the streamwise components of the fluctuating velocity compared to the normal and spanwise components whose predictions by the WALE model are also satisfying. The DSM gives the worst predictions among all models.

\begin{figure}\centering
\includegraphics[height=.30\textwidth]{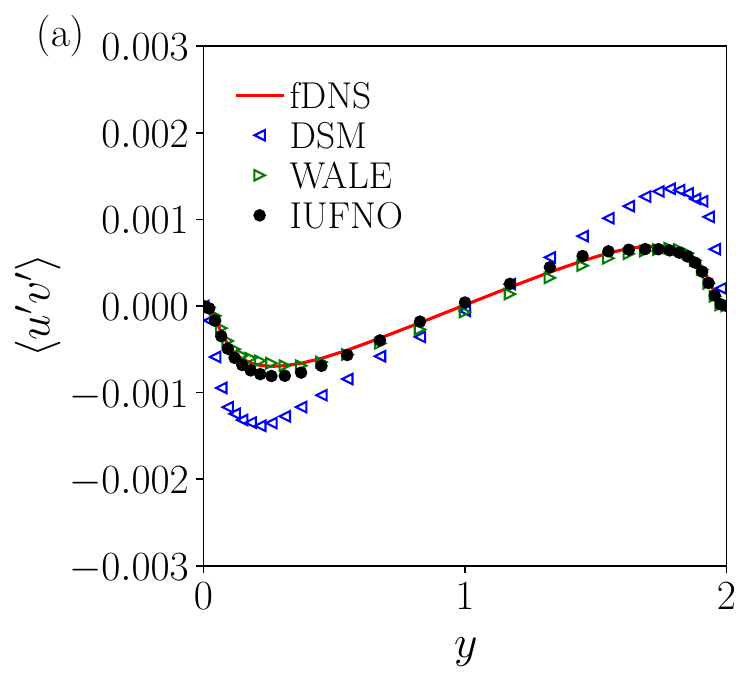}\hspace{-0.10in}
\includegraphics[height=.30\textwidth]{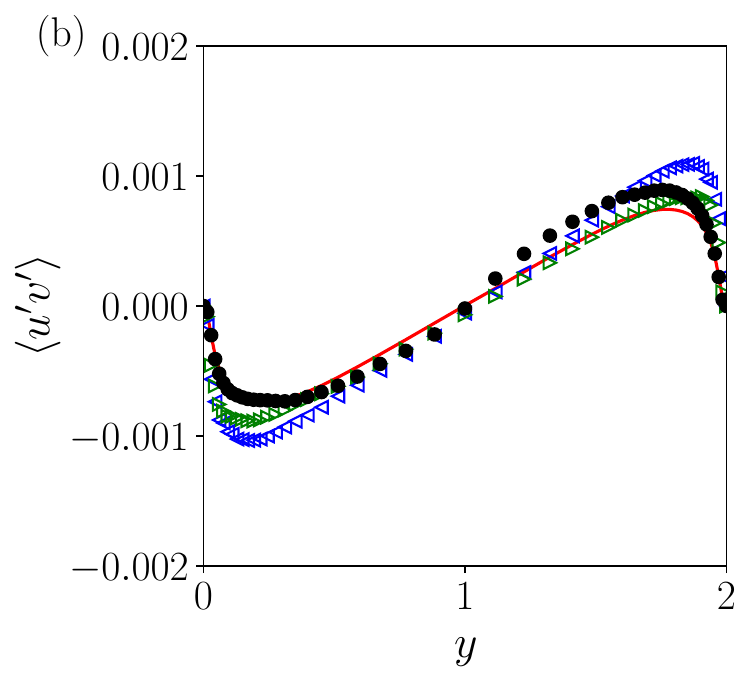}\hspace{-0.10in}
\includegraphics[height=.30\textwidth]{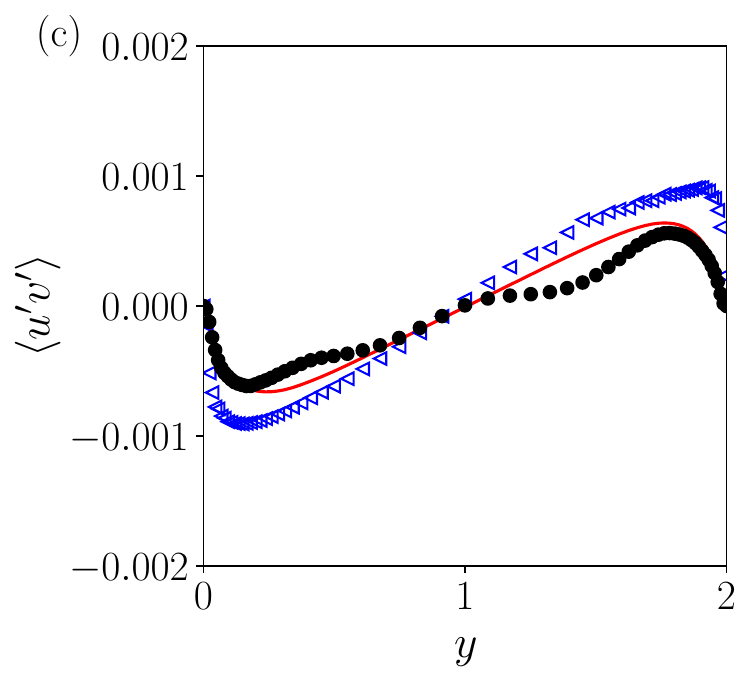}\hspace{-0.10in}
\caption{The variation of the shear Reynolds stress $\langle u'v' \rangle$ at various Reynolds numbers : (a) $Re_{\tau}\approx180$; (b) $Re_{\tau}\approx395$; (c) $Re_{\tau}\approx590$.}\label{fig_Rey_stress}
\end{figure}

The predicted shear Reynolds stresses by different LES models are shown in Fig. \ref{fig_Rey_stress}. The maximum shear Reynolds stresses are located near the upper and lower walls where both the mean shear effects and the velocity fluctuations are strong. The distribution between the two peaks is approximately linear, which is consistent with the literature \cite{Kim1987}. Both the IUFNO and WALE models can predict the Reynolds stress well at $Re_{\tau}\approx180$ and $395$ while apparent discrepancy can be observed for the DSM. At $Re_{\tau}\approx590$, the IUFNO model still performs better than the DSM, but it has some wiggles in the linear region.

\begin{figure}\centering
\includegraphics[height=.33\textwidth]{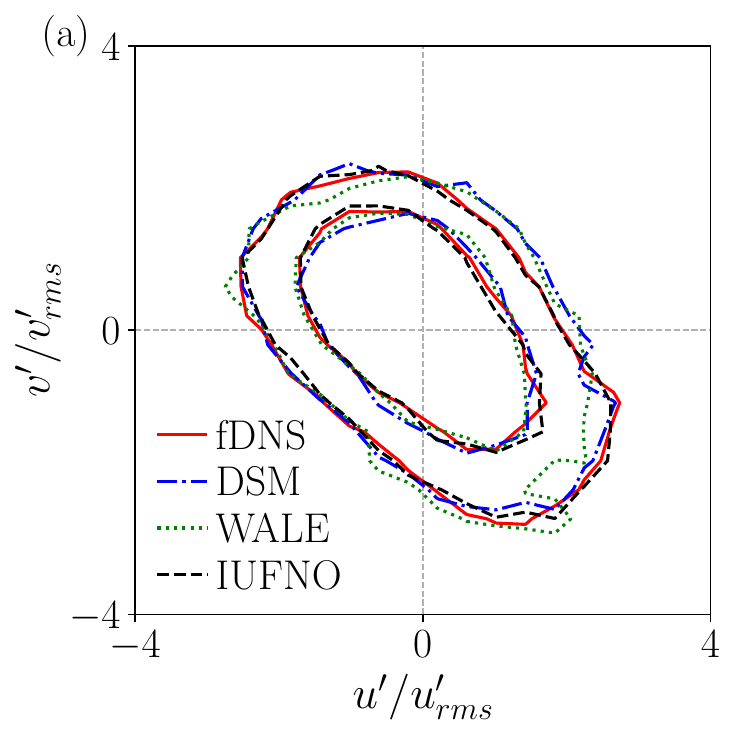}\hspace{-0.10in}
\includegraphics[height=.33\textwidth]{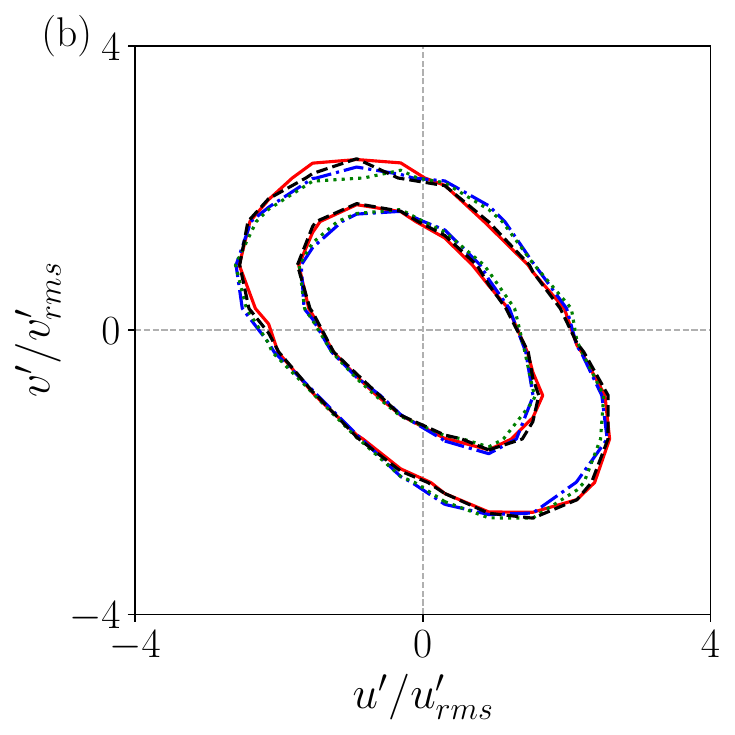}\hspace{-0.10in}
\includegraphics[height=.33\textwidth]{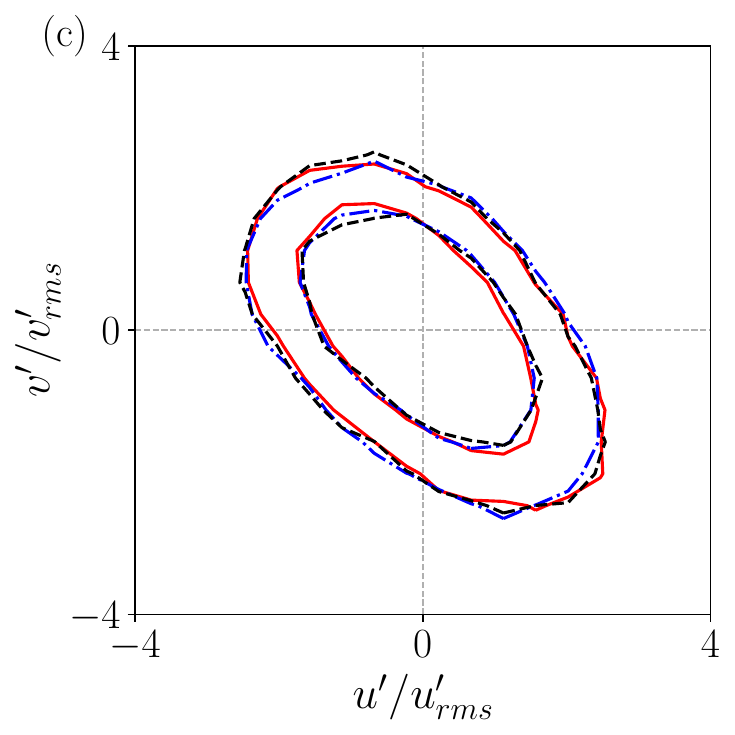}\hspace{-0.10in}

\includegraphics[height=.33\textwidth]{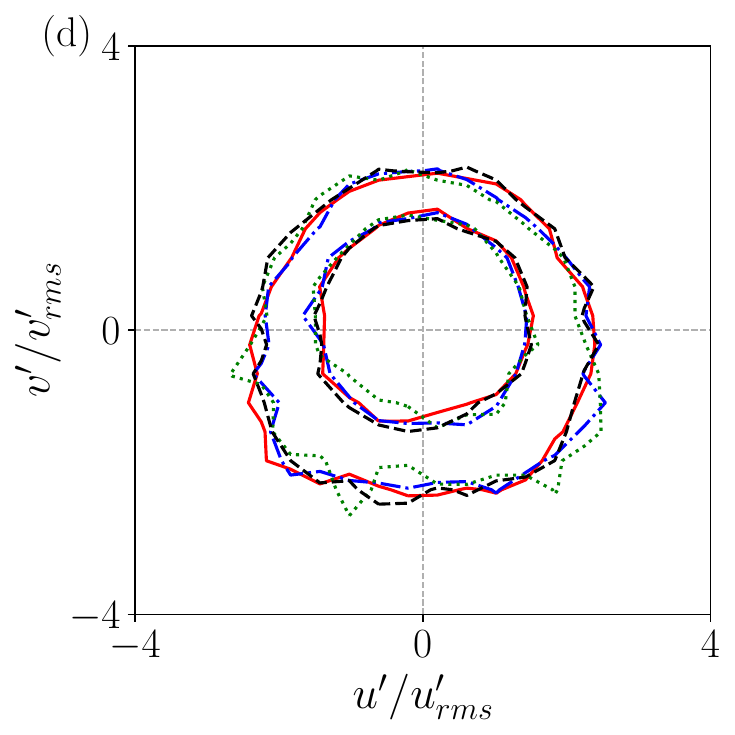}\hspace{-0.10in}
\includegraphics[height=.33\textwidth]{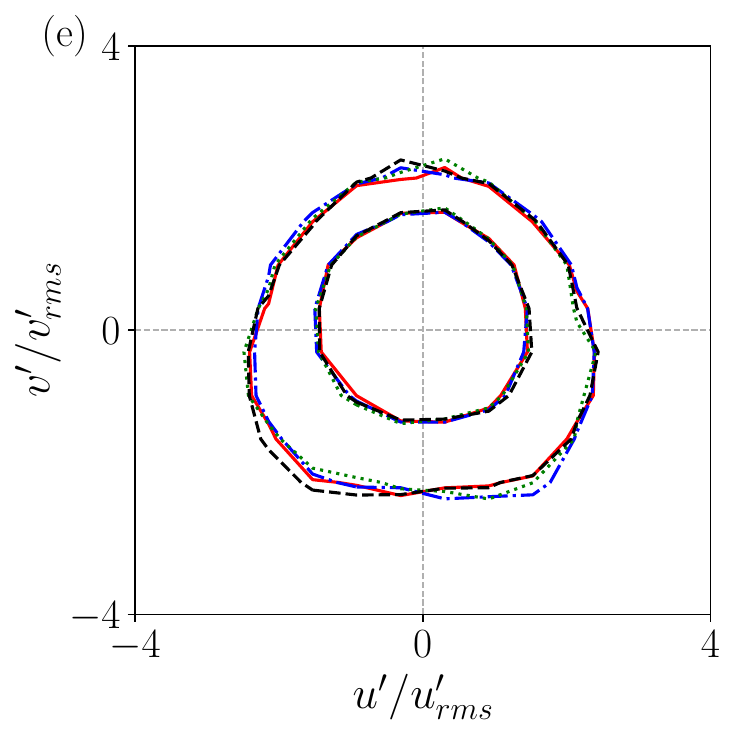}\hspace{-0.10in}
\includegraphics[height=.33\textwidth]{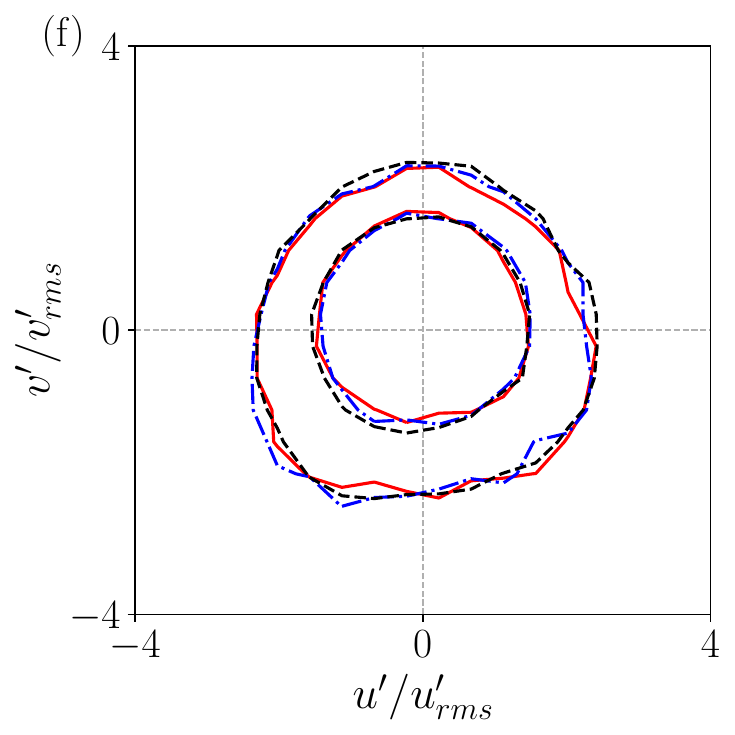}\hspace{-0.10in}
\caption{The joint PDFs of the normalized streamwise and transverse fluctuating velocities at various Reynolds numbers : (a) $Re_{\tau}\approx180$, near the wall ($y/L_y=0.15$); (b) $Re_{\tau}\approx395$, near the wall ($y/L_y=0.12$); (c) $Re_{\tau}\approx590$, near the wall ($y/L_y=0.15$); (d) $Re_{\tau}\approx180$, near the channel center ($y/L_y=0.5$); (e) $Re_{\tau}\approx395$, near the channel center ($y/L_y=0.5$); (f) $Re_{\tau}\approx590$, near the channel center ($y/L_y=0.5$).}\label{fig_Quadrant}
\end{figure}

To further explore behavior of the turbulent shear stress, quadrant analysis is performed \cite{Wallace2016,Xu2021}. We examine the joint PDF of the normalized streamwise and wall-normal fluctuating velocities. The results are displayed in Fig. \ref{fig_Quadrant}. On the basis of quadrant analysis, four events are identified: Q1: $u' > 0, v' > 0$; Q2: $u' < 0, v' > 0$; Q3: $u' < 0, v' < 0$; and Q4: $u' > 0, v' < 0$. Importantly, Q2 represents the ejection event characterized by the rising and breaking up of the near-wall low-speed streaks under the effect of rolling vortex pairs, and Q4 describes the sweeping down of the upper high-speed streaks to the near-wall fluid. Both the ejection and sweep events are gradient-type motions that make the largest contributions to the turbulent shear stress. In contrast, Q1 and Q3 denote the outward and inward interactions that are countergradient-type motions \cite{Wallace2016}.

As shown in Fig. \ref{fig_Quadrant}, at all three Reynolds numbers, the joint PDFs close to the wall are much wider in the second and
fourth quadrants, indicating that shear events (ejection and sweep) are more dominant near the wall rather than the center of the channel. It can be seen that the IUFNO results agree well with the fDNS benchmark. Meanwhile, the predictions by the WALE and DSM models are also satisfying. 

\begin{figure}\centering
\includegraphics[height=.32\textwidth]{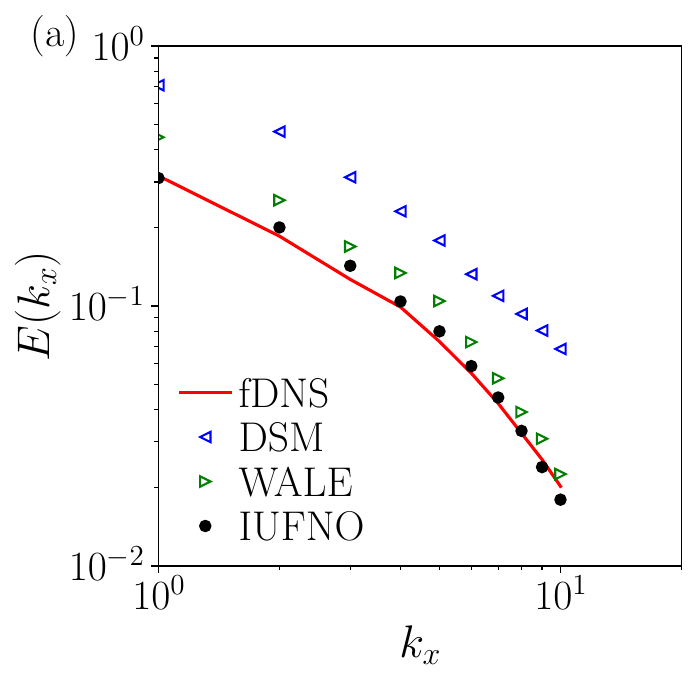}\hspace{-0.08in}
\includegraphics[height=.32\textwidth]{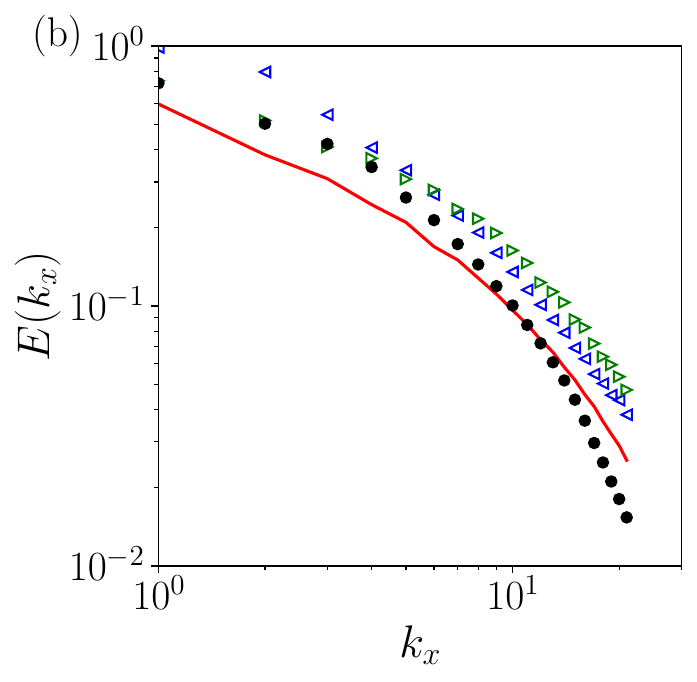}\hspace{-0.08in}
\includegraphics[height=.32\textwidth]{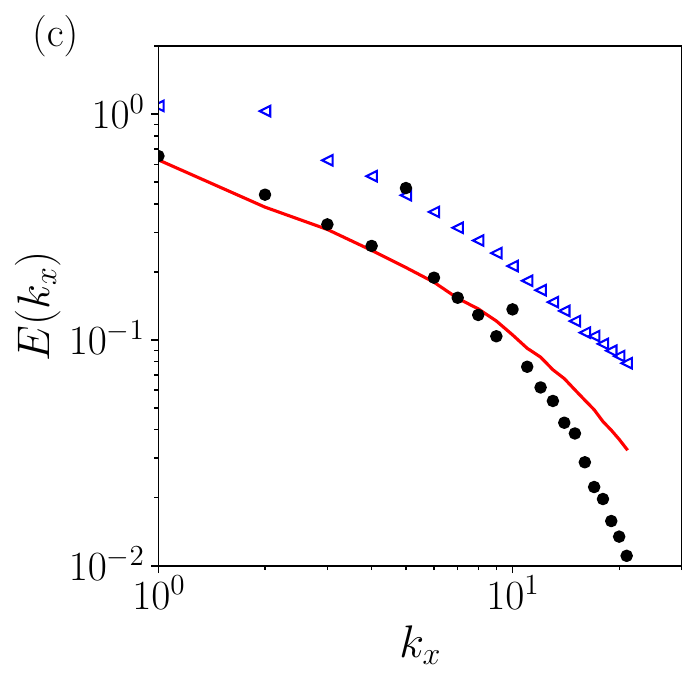}\hspace{-0.08in}

\includegraphics[height=.32\textwidth]{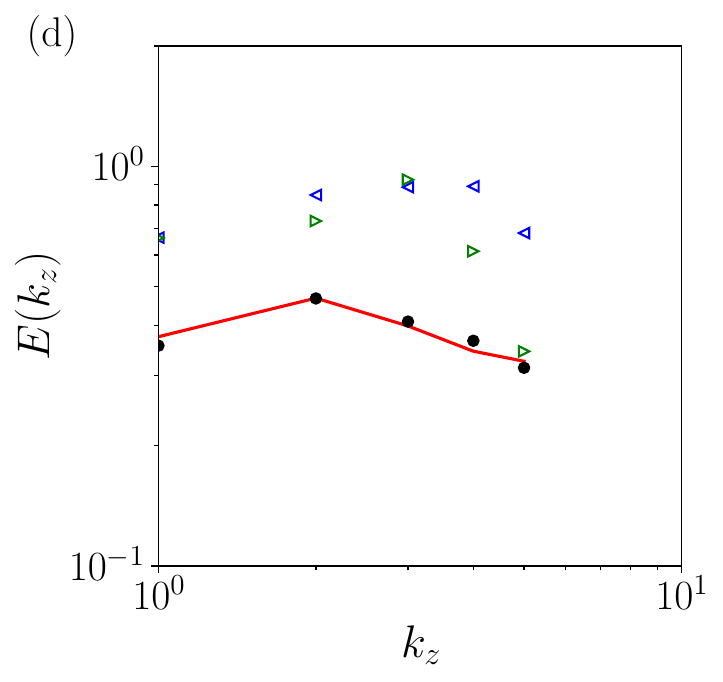}\hspace{-0.08in}
\includegraphics[height=.32\textwidth]{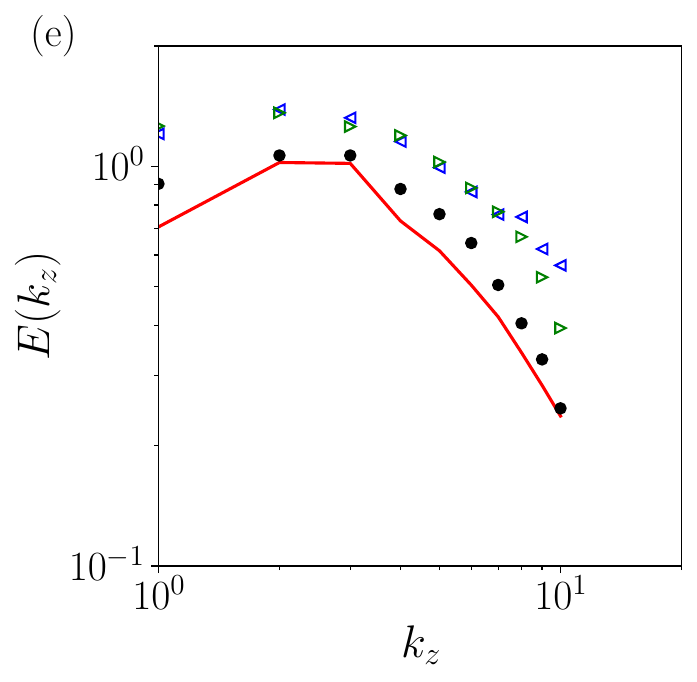}\hspace{-0.08in}
\includegraphics[height=.32\textwidth]{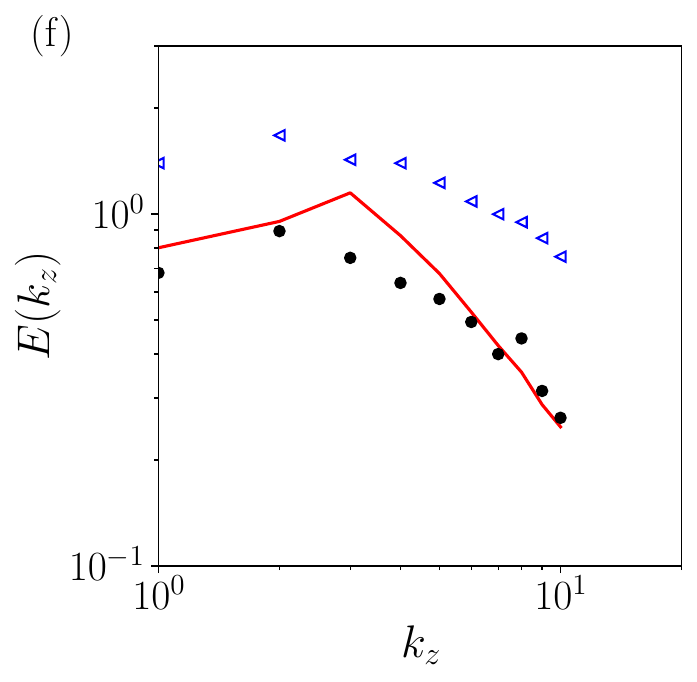}\hspace{-0.08in}

\caption{The streamwise energy spectrum at various Reynolds numbers: (a) streamwise spectrum, $Re_{\tau}\approx180$; (b) streamwise spectrum, $Re_{\tau}\approx395$; (c) streamwise spectrum, $Re_{\tau}\approx590$; (d) spanwise spectrum, $Re_{\tau}\approx180$; (e) spanwise spectrum, $Re_{\tau}\approx395$; (f) spanwise spectrum, $Re_{\tau}\approx590$.}\label{fig_Ek_kx}
\end{figure}

To examine the energy distribution at different scales, we calculate the kinetic energy spectrum in the LES. The streamwise spectra are shown in Figs. \ref{fig_Ek_kx}a, \ref{fig_Ek_kx}b and \ref{fig_Ek_kx}c for $Re_{\tau}\approx180$, $395$ and $590$, respectively. At $Re_{\tau}\approx180$, the predicted spectrum by the IUFNO model agrees well with the fDNS result. Both the DSM and WALE models overestimate the energy spectrum while the WALE model outperforms the DSM. This is also consistent with the results for the rms fluctuating velocity magnitude in Fig. \ref{fig_vel_180}. At $Re_{\tau}\approx395$, the spectrum predicted by IUFNO slightly deviates from the fDNS result, but it still outperforms the DSM and WALE models. For the spectrum at $Re_{\tau}\approx590$, the IUFNO model still outperforms the DSM, however, some nonphysical discontinuities in the IUFNO spectrum can be observed. The cause of such phenomenon is yet unclear.

The spanwise kinetic energy spectra are shown in Figs. \ref{fig_Ek_kx}d, \ref{fig_Ek_kx}e and \ref{fig_Ek_kx}f for $Re_{\tau}\approx180$, $395$ and $590$, respectively. Unlike the streamwise spectrum, the maximum energy in the spanwise spectrum is not contained in the first wavenumber, presumably because there are no dominant largest-scale motions in the spanwise direction. Overall, the IUFNO model has a closer agreement with the fDNS result compared to the WALE and DSM models.

\begin{figure}\centering
\includegraphics[width=.45\textwidth]{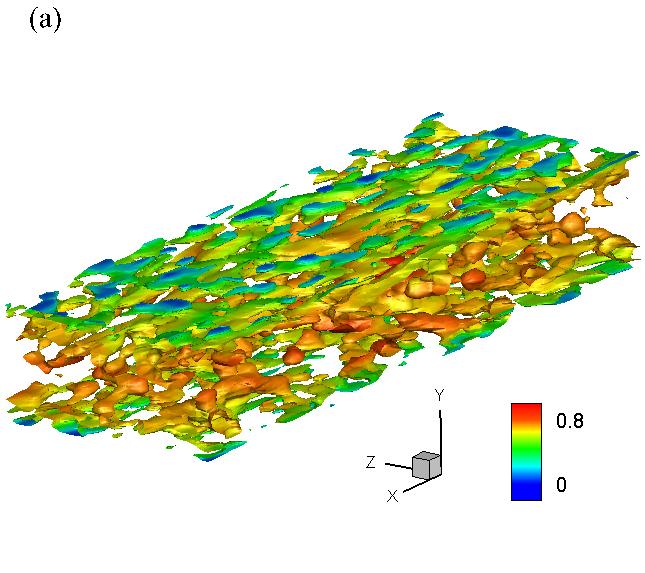}\hspace{-0.05in}
\includegraphics[width=.45\textwidth]{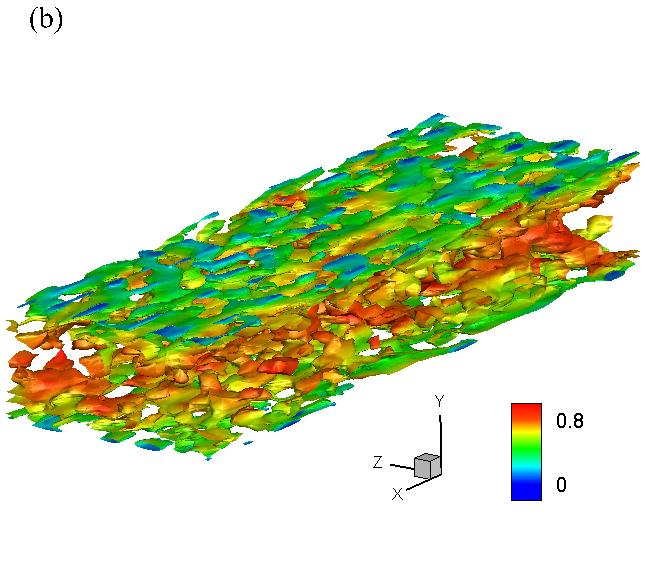}\vspace{-0.05in}

\includegraphics[width=.45\textwidth]{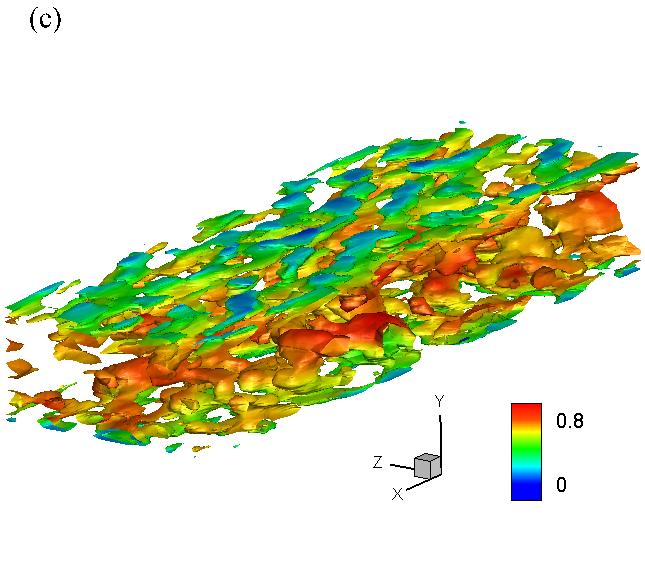}\vspace{-0.05in}
\includegraphics[width=.45\textwidth]{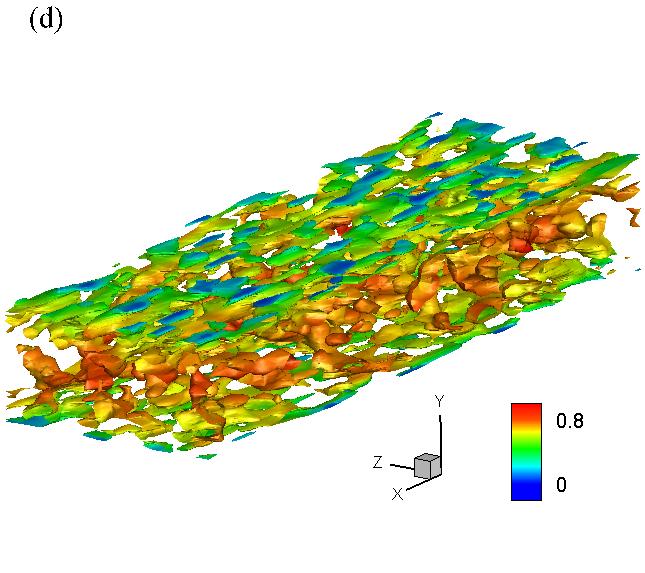}\vspace{-0.05in}

 \caption{The iso-surface of the Q-criterion in the LESs of turbulent channel flow at $Re_{\tau} \approx 180$. Here $Q=0.02$ and the iso-surface is colored by the streamwise velocity: (a) fDNS; (b) DSM; (c) WALE; (d) IUFNO.}\label{fig_Q_180}
\end{figure}

\begin{figure}\centering
\includegraphics[width=.45\textwidth]{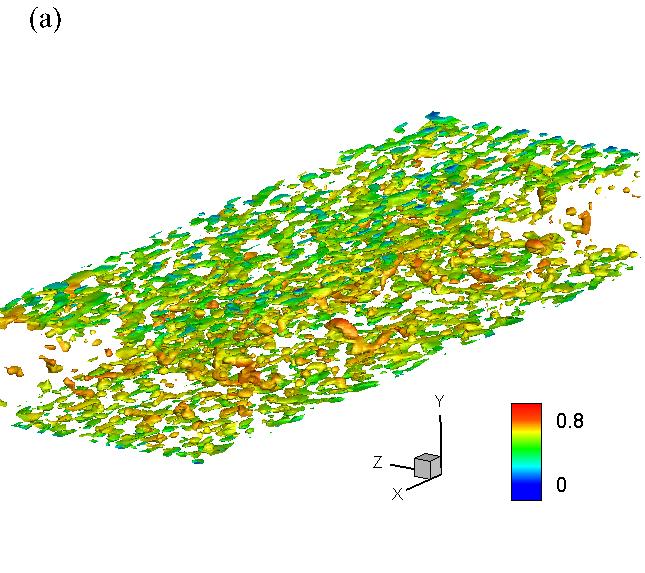}\hspace{-0.05in}
\includegraphics[width=.45\textwidth]{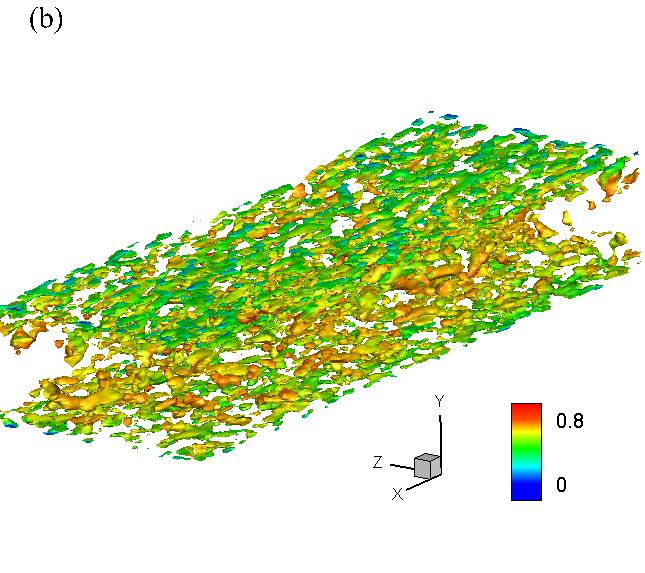}\vspace{-0.05in}

\includegraphics[width=.45\textwidth]{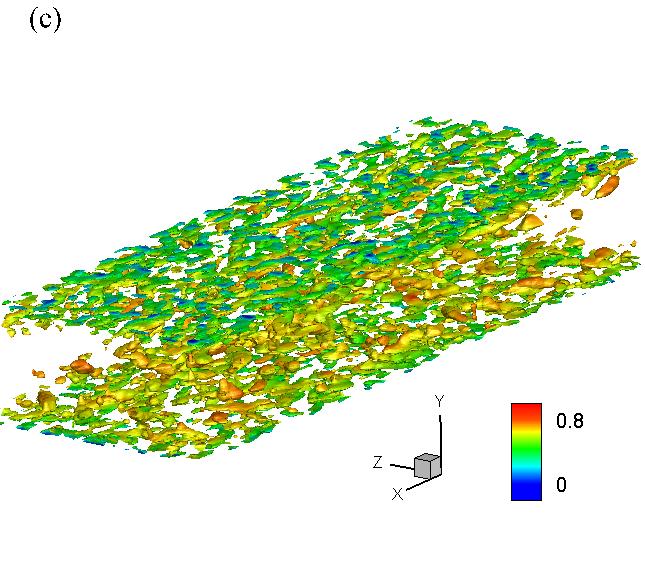}\vspace{-0.05in}
\includegraphics[width=.45\textwidth]{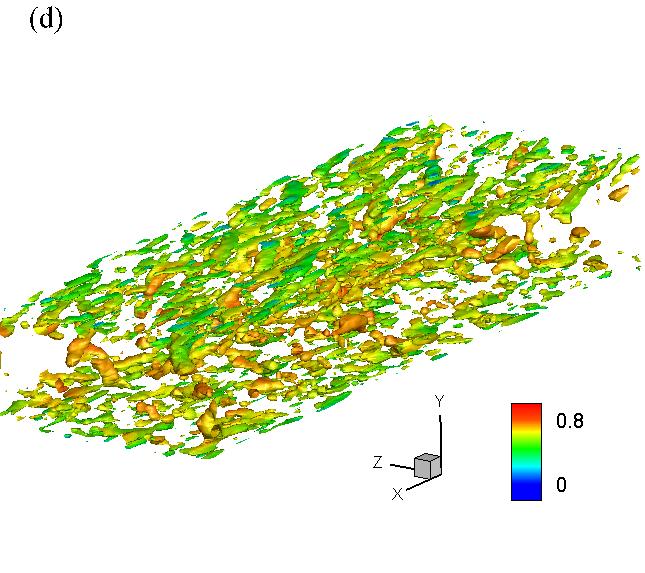}\vspace{-0.05in}

 \caption{The iso-surface of the Q-criterion in the LESs of turbulent channel flow at $Re_{\tau} \approx 395$. Here $Q=0.2$ and the iso-surface is colored by the streamwise velocity: (a) fDNS; (b) DSM; (c) WALE; (d) IUFNO.}\label{fig_Q_395}
\end{figure}

\begin{figure}\centering
\includegraphics[width=.45\textwidth]{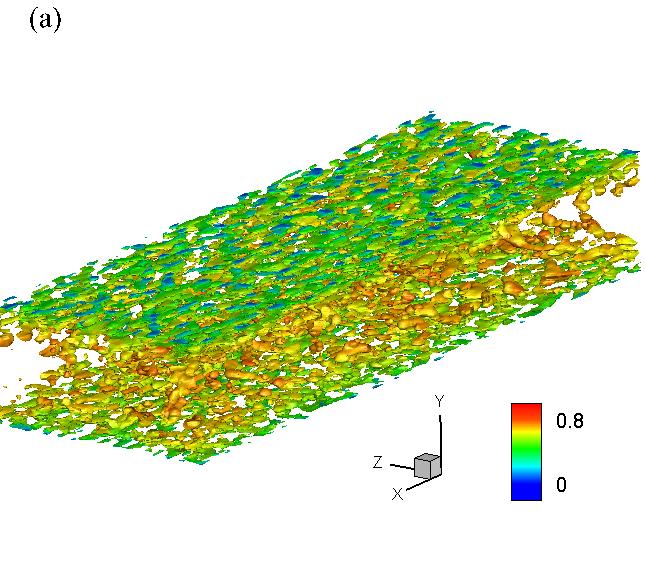}\hspace{-0.05in}
\includegraphics[width=.45\textwidth]{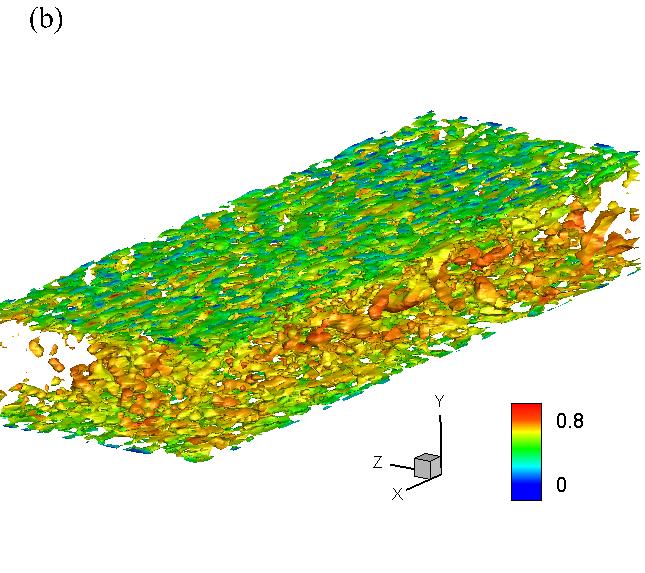}\vspace{-0.05in}

\includegraphics[width=.45\textwidth]{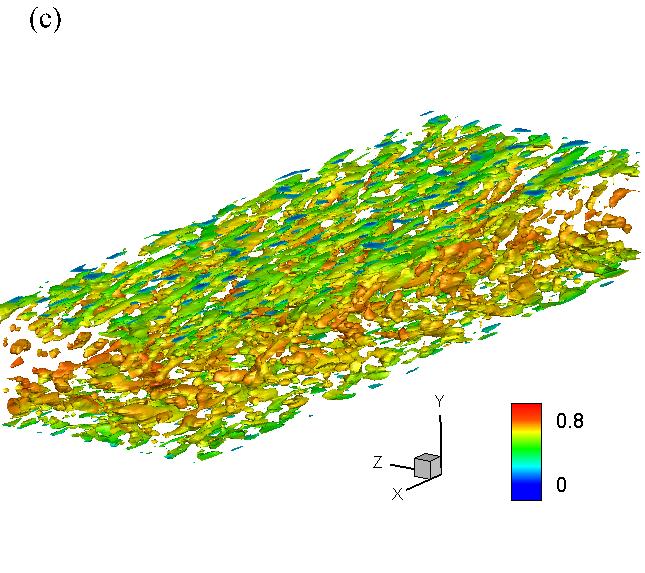}\vspace{-0.05in}

 \caption{The iso-surface of the Q-criterion in the LESs of turbulent channel flow at $Re_{\tau} \approx 590$. Here $Q=0.1$ and the iso-surface is colored by the streamwise velocity: (a) fDNS; (b) DSM; (c) IUFNO.}\label{fig_Q_590}
\end{figure}

To visualize the vortex structure in the turbulent flow field, we examine the Q-criterion defined by \cite{Dubief2000,Zhao2021}
\begin{equation}
Q=\frac{1}{2}[\overline{\Omega}_{ij} \overline{\Omega}_{ij}-(\overline{S}_{ij}-\frac{1}{3}\delta_{ij}\overline{S}_{kk})(\overline{S}_{ij}-\frac{1}{3}\delta_{ij}\overline{S}_{ll})],
\end{equation}
where $\overline{\Omega}_{ij}=\frac{1}{2}(\partial{\overline{u}_{i}}/\partial{x_{j}}-\partial{\overline{u}_{j}}/\partial{x_{i}})$ is the filtered rotation rate. The instantaneous iso-surfaces of $Q$ for $Re_{\tau}\approx180$, $395$ and $590$ are displayed in Figs.~\ref{fig_Q_180}, \ref{fig_Q_395} and \ref{fig_Q_590}, respectively. The iso-surfaces are colored by the streamwise velocities. As the figures depict, the IUFNO models have overall closer agreements with the fDNS results compared to the traditional LES models. Hence, the ability of the IUFNO model to predict the vortex structures is confirmed.

Finally, we compare the computation cost in the LES for different models in Table \ref{tab_cost}. The values in the table are the time consumption (in seconds) for equivalent 10000 DNS time steps. The IUFNO models are tested on the NVIDIA A100 GPU, where the CPU type is AMD EPYC 7763 @2.45GHz. The DSM and WALE simulations are performed on the CPU, which is Intel Xeon Gold 6148 @2.40 GHz. As can be seen in Table \ref{tab_cost}, among the FNO-based models, the original FNO has the lowest computational cost. IUFNO has the largest computational cost as it combines the structures of both the IFNO and UFNO. Nevertheless, the computational efficiency of the IUFNO model is still considerably higher than those of the traditional LES models (i.e. DSM and WALE), considering that the runtimes of the traditional models are not multiplied by the number of adopted computation cores (shown in the parentheses).

\begin{table*}
\begin{center}
\small
\begin{tabular*}{0.8\textwidth}{@{\extracolsep{\fill}}ccccccc}
\hline
$Re_{\tau}$ &DSM &WALE &FNO &IFNO &UFNO &IUFNO \\ \hline
180 &97.4s ($\times$16 cores) &52.1s ($\times$16 cores) &1.3s &2.9s &1.6s &5.0s \\ 
395 &287.2s ($\times$32 cores) &146.3s ($\times$32 cores) &4.5s &9.1s &5.3s &12.1s \\ 
590 &315.3s ($\times$64 cores) &NA &5.9s &11.6s &6.7s &14.9s \\ \hline
\end{tabular*}
\normalsize
\caption{Computational cost of different LES models per 10000 DNS time step.}\label{tab_cost}
\end{center}
\end{table*}

Apparently, the current work shows the potential of IUFNO in the long-term prediction of turbulent channel flow. Nevertheless, it is worth noting that there are still many challenges for machine-learning-based flow predictions, such as high Reynolds number turbulence, turbulent flows with complex geometries, irregular and non-Cartesian grids, etc. On the other hand, it is noticeable that the neural operator-based methods are also quickly evolving and improving these days \cite{Li2022a,Li2023b,Li2023c,Li2023d}. For example, incorporating physical constraints to FNO can lower the required amount of the training data \cite{Li2023d}. The recently proposed geometry-informed neural operator (GINO) can handle complex geometries by combining the FNO with graph neural operator (GNO) which can transform irregular grids into uniform ones \cite{Li2023c}. Meanwhile, the transformer-based frameworks and its combination with FNO have been developed recently, which are also potentially suitable for complex flows \cite{Yang2024,Li2024a}. These are several new aspects that deserve future investigations for complex turbulent flows at high Reynolds numbers.
\section{Conclusions}

In the present study, the Fourier-neural-operator (FNO) and implicit U-Net enhanced FNO (IUFNO) are investigated in the large-eddy simulations (LES) of three-dimensional (3D) turbulent channel flows. In the preliminary test, the IUFNO model outperforms the FNO model in both the training loss and the long-term predictions of velocity statistics. In the \emph{a posteriori} LES tests, the predictive ability of IUFNO is comprehensively examined and compared against the fDNS benchmark as well as the classical DSM and WALE models at various friction Reynolds numbers, namely $Re_{\tau}\approx180$, $395$ and $590$. 

In comparison to the classical DSM and WALE models, the predicted profiles of mean velocity and the fluctuating velocity magnitude by the IUFNO model are closer to the fDNS results even though the performance at the higher Reynolds number is less satisfying. For the probability density functions (PDFs) of the fluctuating velocity, both the IUFNO and WALE models reconstruct well the fDNS results while the DSM results are slightly worse.

Additionally, quadrant analysis is performed by calculating the joint PDF of the normalized streamwise and wall-normal fluctuating velocities. The results indicate that the shear event is quite strong in a region close to the wall while it is very weak in the center of the channel, which is also consistent with the shear Reynolds stress profiles. Overall, the IUFNO model can give quite satisfying predictions for the joint PDFs.  

Further, the streamwise and spanwise kinetic energy spectra are examined. At all three Reynolds numbers, the IUFNO model gives the best predictions. Finally, the Q-criterion is calculated to examine the vortex structures of the turbulent field, and the IUFNO model can predict reasonably well the vortex structures. Considering the relatively low computational cost, these results demonstrate that the IUFNO model is a promising framework for the fast prediction of wall-bounded turbulent flows.

Looking forward to future works, several aspects deserve further investigations, which include augmenting the generalization ability of the model at different Reynolds numbers, the application of the model to more complex wall-bounded flows, and the development and incorporation of wall models in the case of very coarse LES grids.

\begin{acknowledgments}
This work was supported by the National Natural Science Foundation of China (NSFC Grants No. 12302283, No. 12172161, No. 92052301, No. 12161141017, and No. 91952104), by the NSFC Basic Science Center Program (Grant No. 11988102), by the Shenzhen Science and Technology Program (Grant No. KQTD20180411143441009), by Key Special Project for Introduced Talents Team of Southern Marine Science and Engineering Guangdong Laboratory (Guangzhou) (Grant No. GML2019ZD0103), and by Department of Science and Technology of Guangdong Province (Grants No. 2019B21203001, No. 2020B1212030001, and No.2023B1212060001). This work was also supported by Center for Computational Science and Engineering of Southern University of Science and Technology, and by National Center for Applied Mathematics Shenzhen (NCAMS).
\end{acknowledgments}

\end{document}